\documentclass[aps,prb,reprint,superscriptaddress]{revtex4-2}
\usepackage{hyperref}
\usepackage{graphicx}
\usepackage{color}
\usepackage{amsfonts}
\usepackage{amsmath}
\usepackage[utf8]{inputenc}
\usepackage[T1]{fontenc}
\usepackage{mathtools}
\usepackage{bbm}
\usepackage[dvipsnames]{xcolor}
\hypersetup{colorlinks=true, urlcolor=blue, citecolor = blue}

\begin{document}

\title{Scanning gate microscopy of nonretracing electron-hole trajectories\\ in a normal-superconductor junction}

\author{S. Maji}
\email{maji@agh.edu.pl}
\affiliation{AGH University of Krakow, Academic Centre for Materials and Nanotechnology, al. A. Mickiewicza 30, 30-059 Krakow, Poland}

\author{K. Sowa}        
\affiliation{AGH University of Krakow, Faculty of Physics and Applied Computer Science, \\al. A. Mickiewicza 30, 30-059 Krakow, Poland}

\author{M. P. Nowak}        
\email{mpnowak@agh.edu.pl}
\affiliation{AGH University of Krakow, Academic Centre for Materials and Nanotechnology, al. A. Mickiewicza 30, 30-059 Krakow, Poland}

\date{\today}

\begin{abstract}
We theoretically study scanning gate microscopy (SGM) of electron and hole trajectories in a quantum point contact (QPC) embedded in a normal-superconductor (NS) junction. At zero voltage bias, the electrons and holes transported through the QPC form angular lobes and are subject to self-interference, which marks the SGM conductance maps with interference fringes analogously as in normal systems. We predict that for an NS junction at non-zero bias a beating pattern is to occur in the conductance probed with the use of the SGM technique owing to a mismatch of the Fermi wavevectors of electrons and holes. Moreover, the SGM technique exposes a pronounced disturbance in the angular conductance pattern, as the retroreflected hole does not retrace the electron path due to wavevector difference.
\end{abstract}

\maketitle
\section{Introduction}
Electronic transport through a normal-superconductor (NS) interface is governed by the Andreev reflection \cite{Andreev}. It results in conversion of the electron approaching the superconductor into a retroreflected hole and creation of a Cooper pair in the superconductor, provided the electron energy is within the superconducting gap. This elementary process is nowadays a foundation for the functioning of hybrid structures that combine the rich spin physics of semiconductors with the electron pairing provided by the superconductor. Those devices are used for the realization of topological superconductivity \cite{PhysRevLett.105.177002, PhysRevLett.104.040502, Mourik2012}, controllable superconducting \cite{PhysRevLett.115.127001, Casparis2018, PhysRevLett.125.056801} and Andreev qubits \cite{Hays2021}, superconducting diodes \cite{doi:10.1126/sciadv.abo0309}, Andreev molecules \cite{Su2017}, Cooper pair splitters \cite{Wang2022}, and others.

As experiments advance in the creation of devices with complex geometry, such as patterned two-dimensional electron gases (2DEGs) connected to single \cite{Kjaergaard} and multiple superconducting electrodes \cite{Moehle2022, Moehle2021, PhysRevLett.130.116203}, understanding the electronic transport in these devices is of fundamental interest. So far, transport measurements have focused mainly on spectroscopic techniques that lack the ability to determine the spatial properties of quasiparticle propagation. In this work, we propose and theoretically investigate the scanning gate microscopy (SGM) technique as a tool allowing for the visualization of electron and retro-reflected hole paths in a quantum point contact (QPC) embedded in an NS junction.

SGM is a widely used method to visualize electron flow in {\it semiconducting} structures.
It uses a charged atomic force microscope tip that scans above the sample when simultaneously conductance of the system is monitored. In particular, the map of the conductance change can be used to identify the paths of the flowing electrons. This method has been successfully used to visualize the electron flow from a QPC \cite{10.1063/1.1484548}, to attribute the conductance quantization to the occupation of subsequent transverse modes of a QPC \cite{Topinka2000, Topinka} and to demonstrate the coherent electron self-interference \cite{LeRoy, Jura_2007, Electroninterferometer, PhysRevLett.105.166802}. Scanning gate microscopy was also used to image bending of electron trajectories due to an external magnetic field \cite{Aidala2007, Bhandari2016, Bhandari2020} and was theoretically considered in the context of probing the formation of angular lobes and self-interference in 2DEGs \cite{PhysRevB.90.165303, PhysRevB.90.035301, PhysRevB.94.075301, PhysRevB.98.115309} and monoatomic-layered  materials \cite{PhysRevB.96.165310, PhysRevB.95.045304, Mrenca_2015, Prokop_2020}.  

\begin{figure}[h!]
    \centering
    \includegraphics[width=0.98\linewidth]{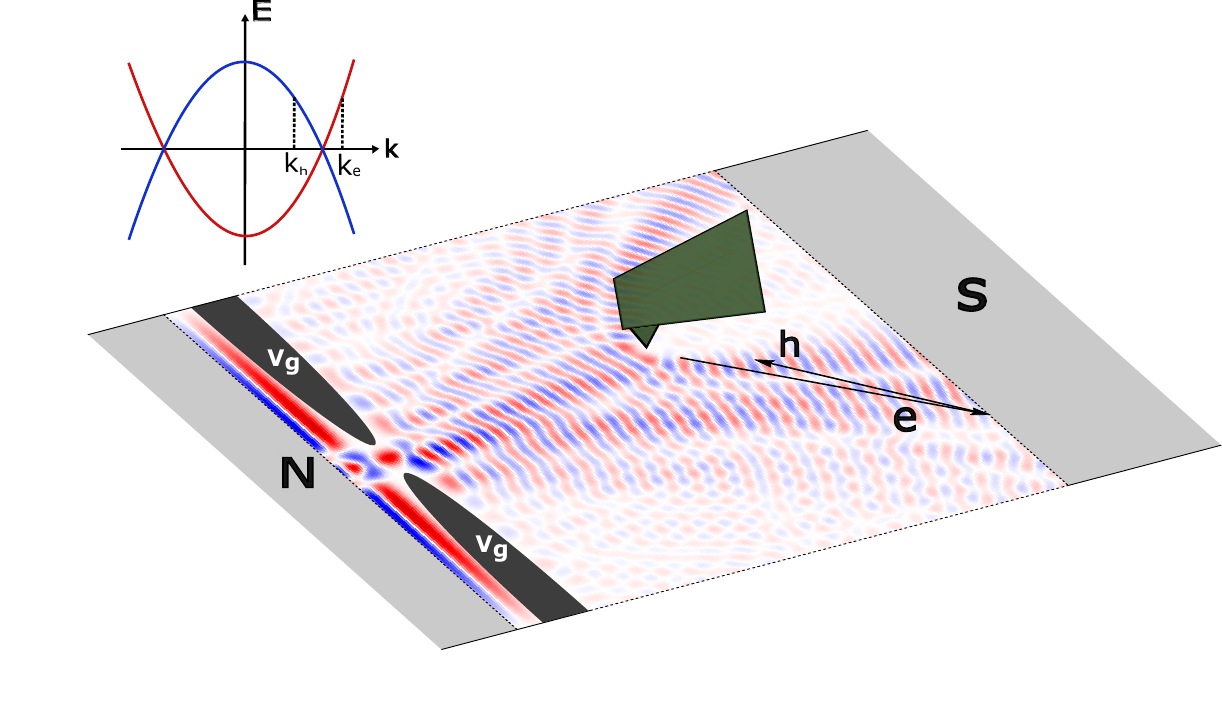}
    \caption{The scheme of the considered system. The QPC (dark grey) is embedded in a NS junction. The normal region, behind the QPC, is scanned by charged SGM tip (dark green) which deflects the electron and hole trajectories in the 2DEG beneath. The self-interference of electrons and holes (blue and red colors show the real part of the electron wave function) leads to oscillations of Andreev enhanced conductance. When the energy of quasiparticles is above the Fermi level the electron and hole have different wavevectors which affects the oscillations and causes the hole not to retrace the electron path (shown with arrows).}
    \label{fig:system}
\end{figure}

In this work, we theoretically investigate the application of the SGM technique in a {\it superconducting} structure: an NS junction that embeds a QPC. In such a system, the hallmark of Andreev reflection is the amplification of conductance \cite{Beenakker, Zhang2017}, which was recently demonstrated in 2DEG QPCs \cite{Kjaergaard2016}. Here, we show that in this system the SGM technique not only reveals the flow in angular lobes of the electrons escaping the QPC, their self-interference, but also interference of retro-reflected holes. Most importantly, this technique unveils the modification of the transport properties of the structure due to the difference in the Fermi velocity of electrons and holes at non-zero bias when the Andreev limit ($\Delta \ll \mu$) is not fulfilled. The latter leads to a pronounced change in the conductance oscillation pattern as the hole does not retrace the electron path. Non-perfect retracing trajectories have been studied so far in the context of electron-hole trajectories in billiards \cite{PhysRevB.73.045324, PhysRevLett.96.237002} and Chladni figures \cite{Libisch2007}. Despite significant progress in studies of superconducting heterostructures, SGM imaging in these systems remains vastly unexplored, with the exception of the experimental demonstration of its use for the visualization of bent electron and hole orbits \cite{Bhandari2020} and the theoretical prediction of probing the supercurrent distribution in Josephson junctions \cite{PhysRevB.106.035432} in the external magnetic field.

\section{Theory}
 \begin{figure}[h!]
    \centering
    \includegraphics[width=0.85\linewidth]{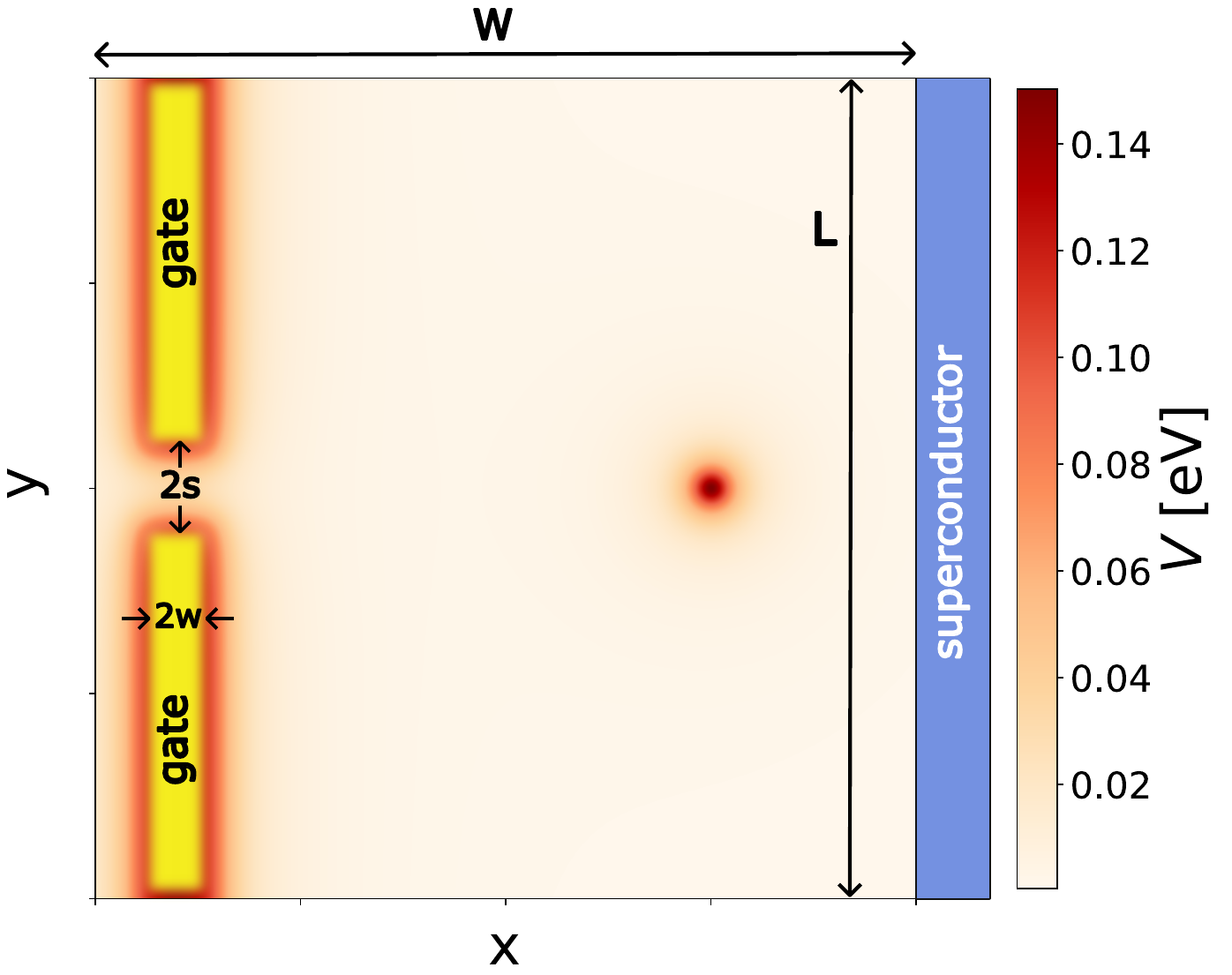}
    \caption{The outline of the system considered in the numerical model. Colors denote the potential distribution in the (normal) scattering region. The superconducting contact is implemented on the right-hand side of the scattering region (blue).}
    \label{fig:supplementary potential}
\end{figure}

The considered system [Fig. \ref{fig:system}] consists of a QPC embedded in the normal region of the NS junction. We consider the zero-magnetic field case, hence we use a spinless Hamiltonian written in the electron-hole basis
\begin{multline}
\label{eqn:Hamiltonian}
H = (\hbar^2\textbf{k}^2/2m^*+ V_{\mathrm{QPC}}(x,y) + \\ V_{\mathrm{SGM}}(x,y, x_{\mathrm{tip}}, y_{\mathrm{tip}}) -\mu)\tau_z + \Delta(x)\tau_x,
\end{multline}
 acting on the wave-function $\Psi = (\Psi_e, \Psi_h)^T$. $\tau_x$ and $\tau_z$ are the $x$ and $z$ Pauli matrices and $\mu = 10$ meV is the chemical potential.
 $V_\mathrm{QPC}(x,y)$ is the QPC potential modeled after Ref. \cite{Davies} as 
\begin{multline}
V_\mathrm{QPC}(x,y)/(-eV_g)=\frac{1}{\pi}[\arctan((w+x-x_1)/d)+\\ \arctan((w-x+x_1)/d)]\\-g(s+y,w+x-x_1)-g(s+y,w-x+x_1)\\-g(s-y,w+x-x_1)-g(s-y,w-x+x_1),
\end{multline}
where 
\begin{equation}
    g(u,v)=\frac{1}{2\pi}\arctan(uv/dR),
\end{equation}
and $R=\sqrt{u^2+v^2+d^2}$. $w$ is the width of the QPC gates, $s$ is their separation, and $d$ is the distance between the gates and the 2DEG [see Fig. \ref{fig:supplementary potential}]. We take $w = 100$ nm, $s = 75$ nm and $d = 50$ nm compatible with the experimentally implemented QPCs \cite{LeRoy, Aoki, Iagallo, Jura_2007} and place the QPC at $x_1 = 200$ nm.
 
To model the SGM potential, we use a potential distribution \cite{Szafran} 
\begin{equation}
V_{\mathrm{SGM}}(x,y)=\frac{V_{\mathrm{tip}}}{1+\frac{(x-x_{\mathrm{tip}})^2+(y-y_{\mathrm{tip}})^2}{d_{\mathrm{sgm}}^2}}
\end{equation}
where, we set $V_\mathrm{tip}=0.05$ eV and $d_{\mathrm{sgm}}=50$ nm. The position of the SGM tip is determined by the pair of coordinates ($x_{\mathrm{tip}},y_{\mathrm{tip}}$). We assume that the scattering region is connected to the biased normal lead just before the QPC (see gray $N$ region in Fig. \ref{fig:system}), which serves as the source of incoming electrons, and to a grounded superconducting lead, which provides the Andreev reflection. 

The zero-temperature conductance at bias $V_b = -E/|e|$ is calculated according to the Landauer-Buttiker formula \cite{linear_note} 
\begin{equation}
G(E)= \frac{2e^2}{h} \cdot (N(E)-R_{ee}(E)+R_{he}(E)).
\end{equation} 
$N$ is the number of electronic modes in the normal lead, $R_{ee}(E)$ electron-to-electron reflection coefficient, and $R_{he}(E)$ electron-to-hole reflection coefficient \cite{PhysRevB.97.045421}. The coefficients $R_{ee}(E)$ and $R_{he}(E)$ are obtained from the scattering matrix of the system calculated at the energy $E$,
\begin{equation}
    R_{\alpha,\beta}(E) = \mathrm{Tr}\left( [S_{\alpha,\beta}(E)]^\dagger S_{\alpha,\beta}(E)\right),
\end{equation}
where $S_{\alpha,\beta}(E)$ is the scattering matrix block corresponding to the particles of type $\beta$ injected in the normal lead and scattered back as particle type $\alpha$ into the same lead. To account for the large size of the device in the direction parallel to the QPC gates, we apply open boundary conditions at the edges of our system in the $y$ direction. For concreteness, we assume InSb material parameters $m^*=0.014m$ and the superconducting gap of $\Delta = 2$ meV corresponding to superconductors such as Nb, NbTiN \cite{Salimian, Zhi, PhysRevB.99.245302, Satchell}. We set the system size to $W = 4000$ nm, $L = 4000$ nm. We discretize the Hamiltonian Eq. (\ref{eqn:Hamiltonian}) on a square mesh with the lattice constant $a = 10$ nm and solve the transport problem using the scattering matrix approach implemented in the Kwant package \cite{Christoph}. The data for the plots were obtained using the Adaptive package \cite{Nijholt2019} and the code used for the simulations is available in an open repository \cite{zenodo_repository}.

\section{Results}
\begin{figure}[h!]
    \centering
    \includegraphics[width=0.85\linewidth]{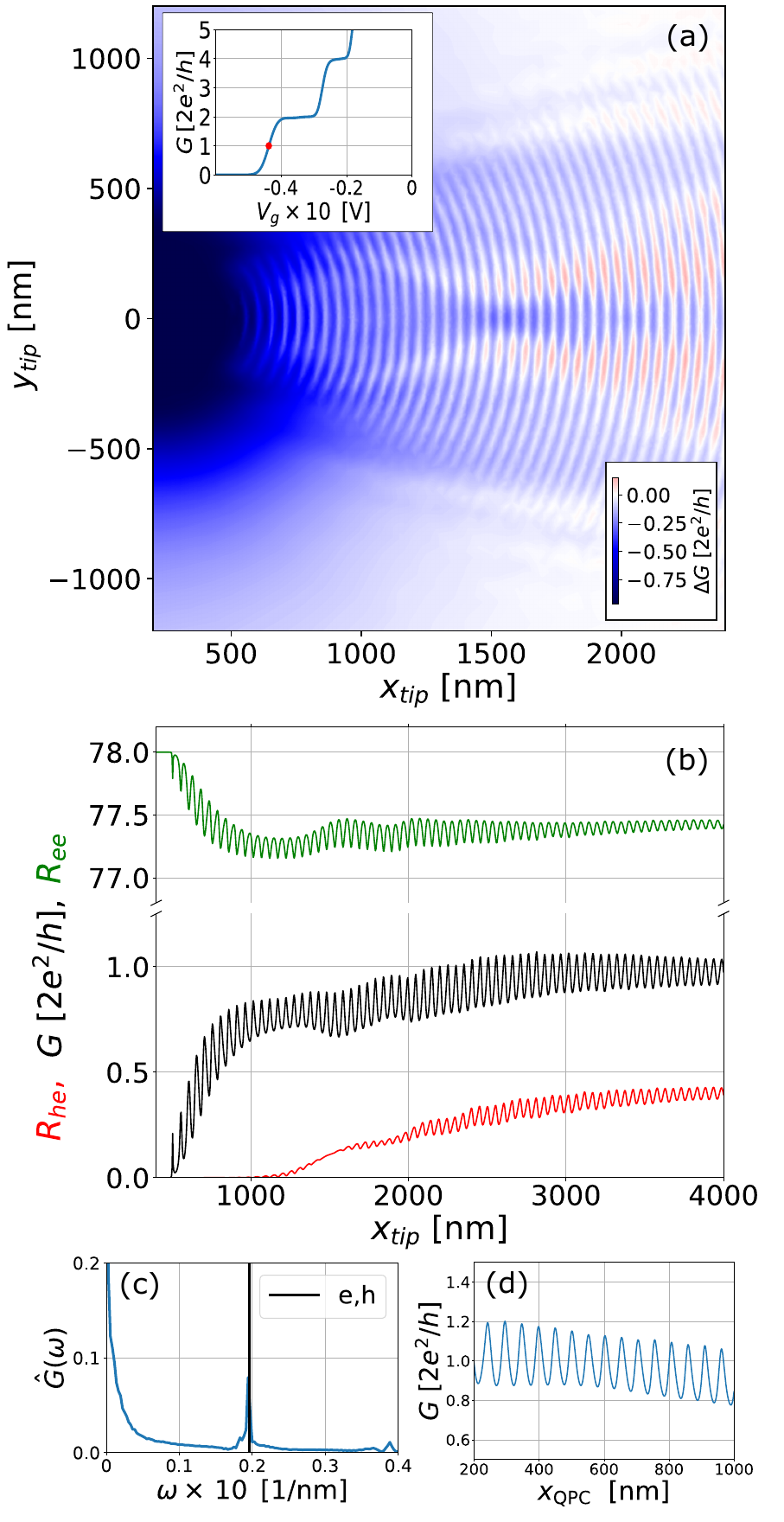}
    \caption{(a) The map of the change of the conductance introduced by the SGM tip at zero bias voltage with a single lobe flow visible and periodic fringes due to electron and hole self interference. The inset shows conductance versus the QPC gate voltage with the red point denoting the $V_g$ value chosen for the calculation of the map. (b) Conductance (black), electron-electron (green) and electron-hole (red) transmission coefficients obtained for $y_{\mathrm{tip}} = 0$. (c) The Fourier transform of the conductance from (b). (d) Conductance versus the position of the QPC with the SGM tip set in constant distance from the superconductor interface.}
    \label{fig:zero energy}
\end{figure}

The electron incident from the normal lead propagates through the QPC, whose potential is controlled by the $V_g$ voltage; then Andreev reflects at the superconducting contact; and finally, the resulting hole is scattered back to the original normal lead. 
The inset of Fig. \ref{fig:zero energy}(a) shows a typical conductance curve versus gate voltage \cite{QPC_note}, where conductance is amplified by the Andreev reflection and quantized with plateaus in multiples of $4e^2/h$. For the following calculations, we choose the value $V_g$ so that the resulting conductance is $2e^2/h$---at the first step---when a single lobe in the SGM map is expected \cite{Electroninterferometer}.

In the map of Fig. \ref{fig:zero energy}(a) we show the change of the conductance (difference between the conductance obtained in the presence of the tip and the conductance obtained in the absence of the tip) when the system is scanned by the SGM tip. On the map, pronounced radial fringes due to self-interference of the quasiparticles are present.

The conductance cross-section for the SGM $y$ coordinate set in the middle of the QPC constriction ($y_{\mathrm{tip}}=0$) is shown in Fig. \ref{fig:zero energy}(b) with the black curve. When the tip is moved away from the QPC constriction, the conductance rapidly increases as the electron flow through the QPC is unblocked and then exhibits periodic oscillations. The oscillation fringes are separated by half of the Fermi wavelength due to interference between the waves reflected by the QPC itself and those reflected back by the tip \cite{Electroninterferometer}. In Fig. \ref{fig:zero energy}(c) we show the Fourier transform of the conductance. We observe that the oscillations occur with a single period that corresponds to the Fermi wavevector of electrons and holes $k_{e/h} = \sqrt{2m^*\mu}/\hbar$ \cite{k_note} denoted by the black vertical line in the inset with $\omega_{e/h}={1}/{(\lambda_\mathrm{e/h}\times10^9)}$ and $\lambda_{e/h}=\pi/k_\mathrm{e/h}$. In a normal system such oscillations are a signature of constructive and destructive interference of the electronic wave that occurs between the QPC constriction and the SGM tip. In fact, by changing the distance between the QPC and the SGM tip, we observe conductance oscillations [Fig. \ref{fig:zero energy}(d)] that confirm that a similar process also takes place here. We have also checked the conductance oscillation when the superconducting interface was moved away from the QPC and SGM tip and observed their negligible amplitude (not shown). This allows us to conclude that the main path of the interference is between the QPC and the SGM tip.

In Fig. \ref{fig:zero energy}(b) we show the transmission coefficients of the electron-electron (green curve) and electron-hole (red curve) transport process. The values of the $R_{ee}$ coefficient are high and result from the wide lead considered at the left boundary (which results in the number of current-carrying modes $N = 78$), but at the same time most of the modes are reflected from the nearly closed QPC, which conducts only one mode, resulting in $N - R_{ee} \simeq 1$. $R_{he}$ has a considerable magnitude which means that the injected electron, after passing the QPC and scattering at the SGM tip potential, is converted into a hole at the superconducting interface that is traced back to the normal electrode. Note that in this case the interference of electrons and holes is indistinguishable, as they both have the same wavevector.

\begin{figure}[h!]
    \centering
    \includegraphics[width=0.85\linewidth]{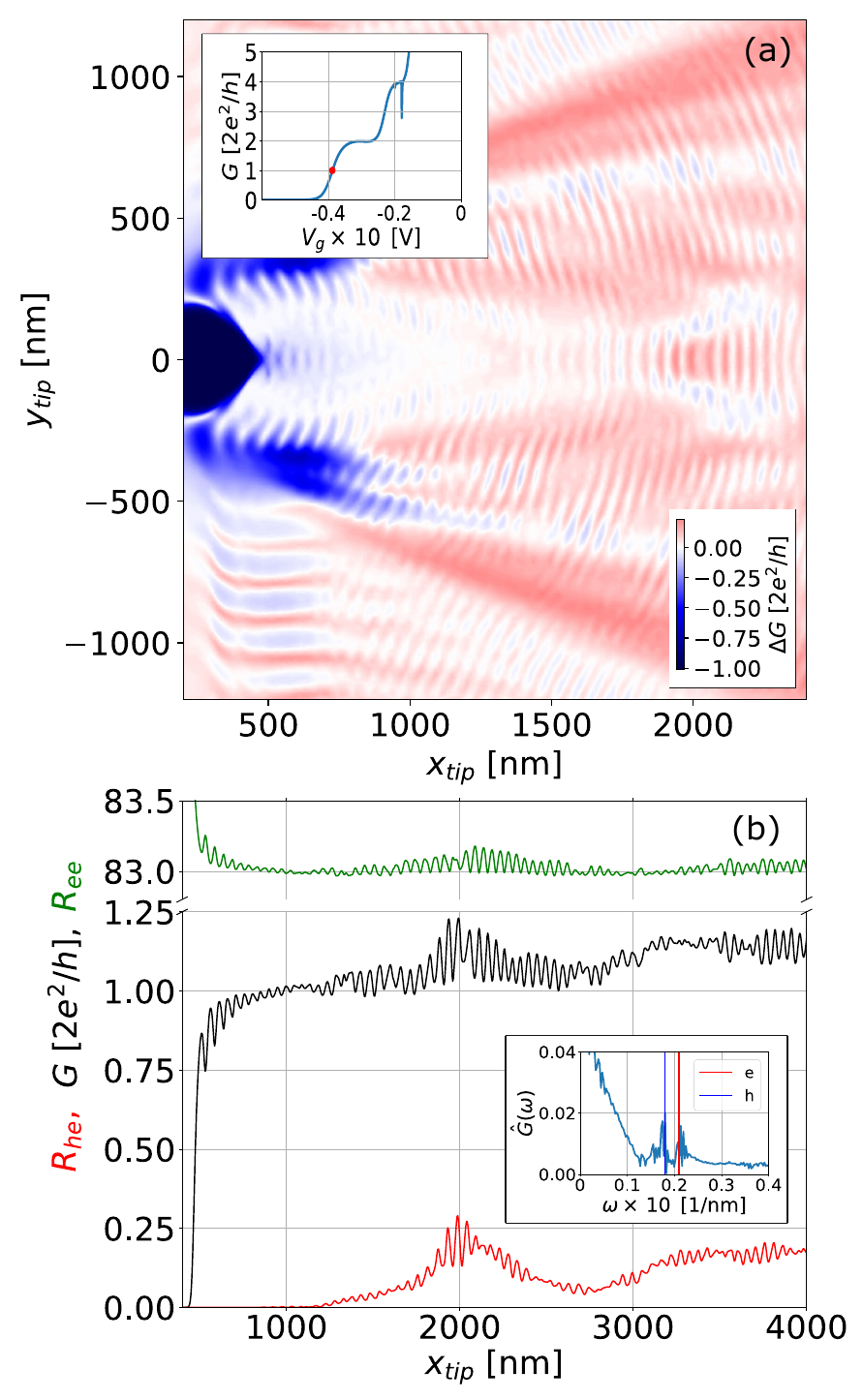}
    \caption{(a) The map of the change of the conductance introduced by the SGM tip at non-zero voltage ($V_b = -0.75\Delta/|e|$) with a disturbed fringes pattern due to the non-retracing hole trajectories. The inset shows conductance versus the QPC gate voltage with the red point denoting the $V_g$ value chosen for the calculation of the map. (b) Conductance (black), electron-electron (green) and electron-hole (red) transmission coefficients obtained for $y_{\mathrm{tip}} = 0$. The inset shows the Fourier transform of the conductance cross-section with two prominent frequencies for electrons and holes.}
    \label{fig:non-zero energy}
\end{figure}

Let us now move to the non-zero voltage bias case. We consider $V_b = -0.75\Delta/|e|$ such that the electron and hole attain considerably different wavevectors, but the bias voltage is still not high enough to induce smearing in the conductance fringes due to increased scattering \cite{leroy2002imaging}. We again fix $V_g$, so the conductance without the tip is in the middle of the first conductance step (see the inset of Fig. \ref{fig:non-zero energy}(a)). The corresponding $\Delta G$ map is shown in Fig. \ref{fig:non-zero energy}(a), where we observe a significant modification in the conductance pattern as compared to the case of Fig. \ref{fig:zero energy}(a), with three main ingredients: i) change of the pattern of oscillation fringes; ii) resonant features disturbing the previously clear single-lobe flow; iii) general amplification of the conductance by the tip. As we will show in the following, they all are a signature of unequal electron and hole wave vectors.

\begin{figure*}[ht!]
    \centering
    \includegraphics[width=0.9\linewidth]{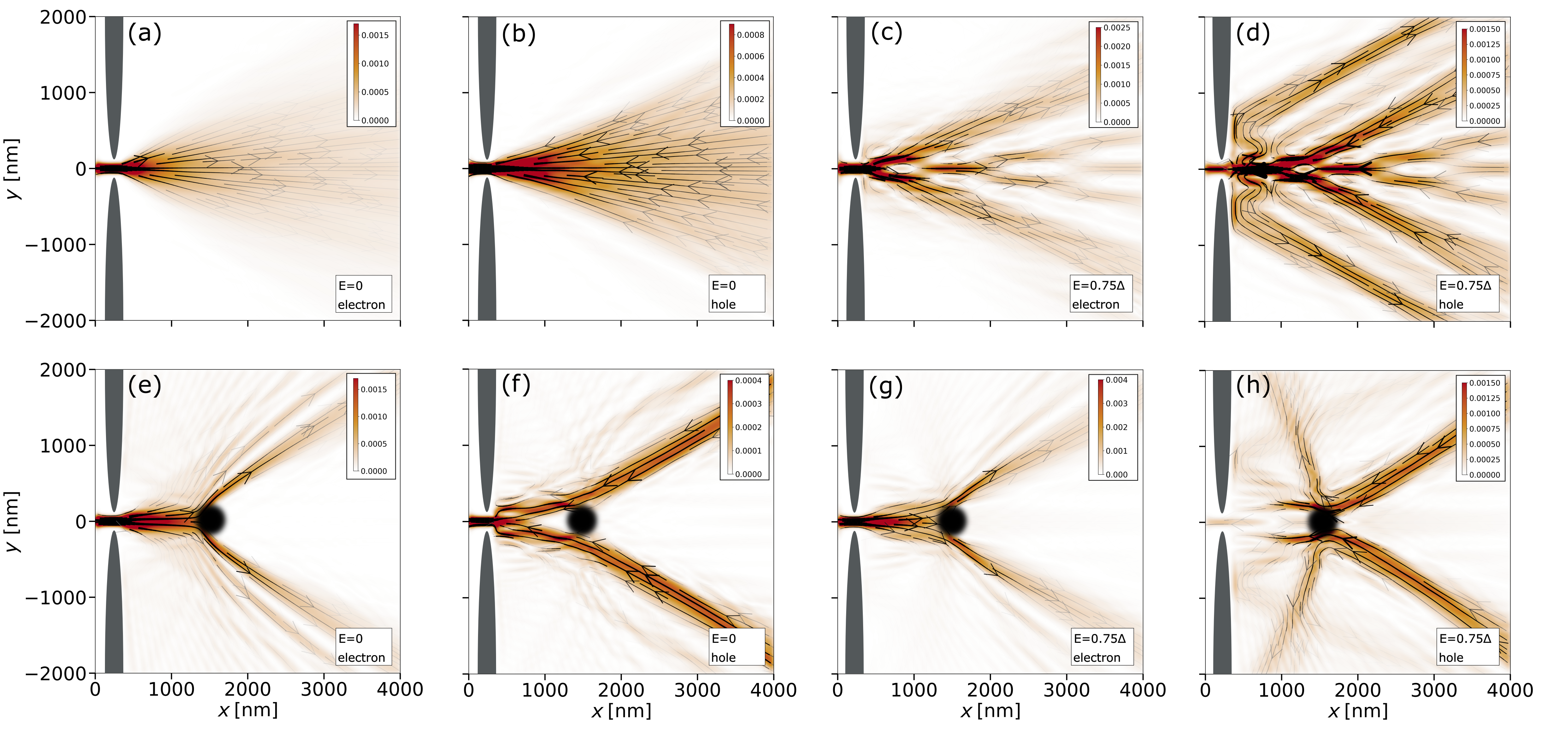}
    \caption{(a)(c)(e)(g) Electron and (b)(d)(f)(h) hole probability currents obtained at (a)(b)(e)(f) zero-energy and at (c)(d)(g)(h) non-zero energy. The upper row shows the results without the SGM tip, and the lower panel presents the results with the SGM tip included.}
    \label{fig:currents}
\end{figure*}

The conductance cross-section is shown in Fig. \ref{fig:non-zero energy}(b) with the black curve. We observe that the oscillations are disturbed by a beating pattern and correspondingly in the Fourier transform (the inset of Fig. \ref{fig:non-zero energy}(b)) we find two leading frequencies corresponding to different values of the electron and hole wavevector ($k_e$ and $k_h$) obtained as $k_{e/h} = \sqrt{2m^*(\mu \pm E)}/\hbar$ with $+$ for the electron and $-$ for the hole (see the schematic dispersion relation in Fig. \ref{fig:system}). This allows for distinguishing the electron and hole contributions to the conductance oscillations.

Inspecting the electron and hole contributions to the conductance shown with green and red curves in Fig. \ref{fig:non-zero energy}(b), respectively, we observe a significant reduction of the $R_{ee}$ coefficient compared to the zero-energy case of Fig. \ref{fig:zero energy}(b). This means that at the first conductance step the probability of the electron that has passed the QPC to go back to the source electrode is negligible. This phenomenon results in a small value of $R_{he}$ coefficient. This means that the almost unity conductance is obtained mainly because of the lack of electron reflection to the left lead and not because of the Andreev reflected hole scattered back there. 

This initially puzzling result becomes clear when one inspects the probability current maps shown in Fig. \ref{fig:currents}. First, when considering the zero-energy case we see that the electron (a) and (b) the hole currents are mostly the same, as the electron injected from the QPC is Andreev reflected and the resulting hole is back-focused into the QPC constriction. Upon introduction of the SGM tip (marked as a gray circle in the bottom row of Fig. \ref{fig:currents}) we see a deflection of the electron trajectory, but the hole does trace the electron path. 

The situation is strikingly different in the non-zero energy case. First, in the absence of the tip, we clearly observe the creation of resonant features that perturb the electron [Fig. \ref{fig:currents}(c)] and hole [Fig. \ref{fig:currents}(d)] flow. The electron approaches the superconductor interface with the wavevector $k_e$ at an angle $\varphi_e$ defined with respect to the normal to the interface. After the Andreev reflection, the hole leaving the interface has the wavevector $k_h \ne k_e$. The momentum component along the interface is preserved and, hence, for non-zero $E$ the normal components of $k_e$ and $k_h$ are different. As a result, the hole returns from the interface following a trajectory determined by $\varphi_h \ne \varphi_e$ and therefore does not retrace the electron path \cite{PhysRevB.73.045324} [see the arrows in Fig. \ref{fig:system}]. As a result the Andreev reflected hole does not fully focus on QPC constriction, which results in a low value of the $R_{he}$ coefficient. In Fig. \ref{fig:currents}(d) we also observe streams of hole current that point {\it from} the QPC to the sink electrodes located at the top and bottom of the system. The introduction of the SGM tip at those locations results in an amplification of the conductance, as the hole reflected by the tip to the QPC is now able to amplify the conductance, leading to an increase of the conductance by the tip shown in Fig. \ref{fig:non-zero energy}(a). Finally, upon introduction of the tip in Figs. \ref{fig:currents}(g) and (h), we clearly observe the effect of different incident and reflection angles, which deflects the hole current from the initial trajectory of the incident electron and, in turn, causes disturbance of the conductance pattern as seen in the map of Fig. \ref{fig:non-zero energy}(a).

\begin{figure}
    \centering
    \includegraphics[width=0.75\linewidth]{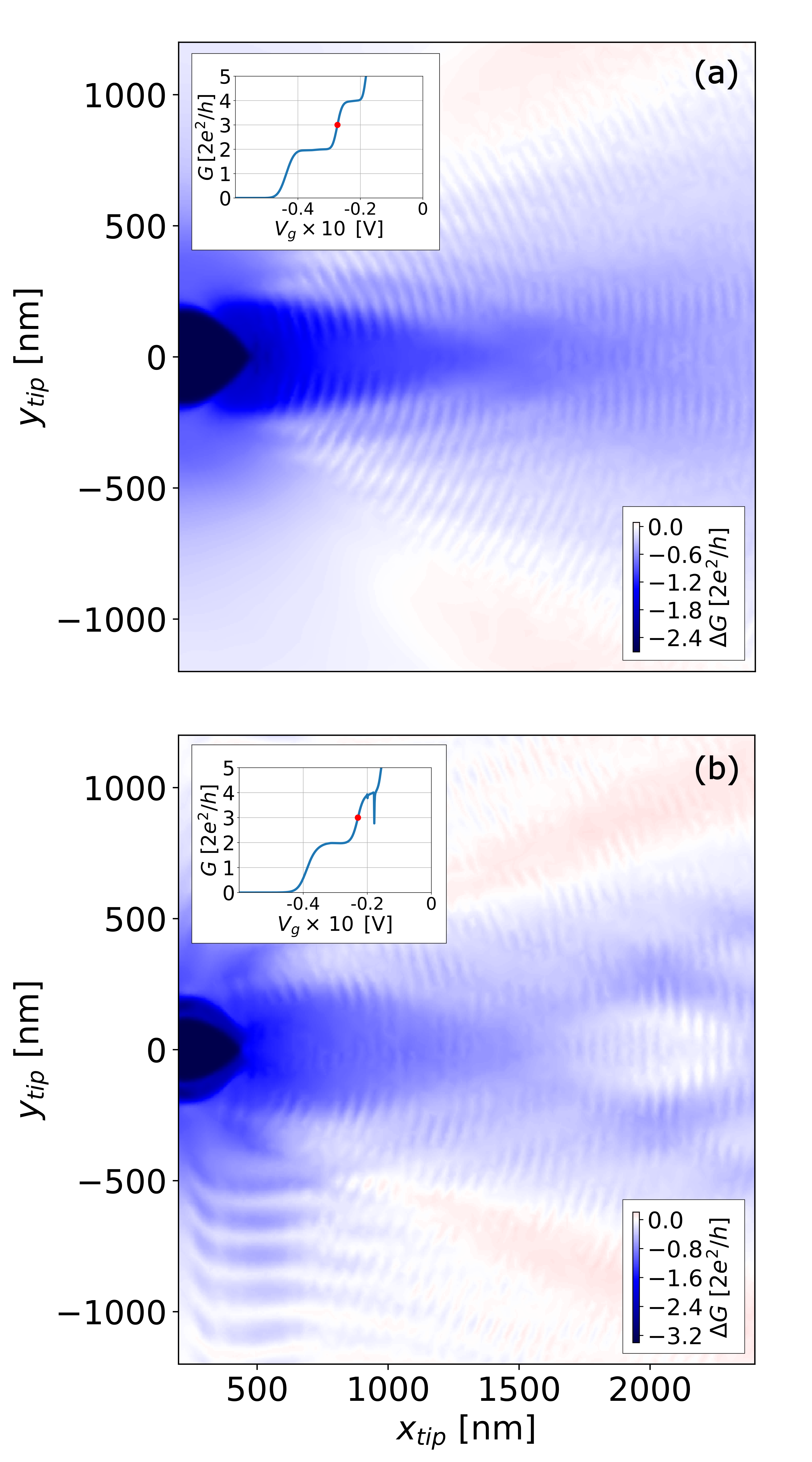}
    \caption{SGM conductance maps obtained for (a) $E = 0$, (b) $E = 0.75 \Delta$. The insets show QPC conductance versus gate voltage with the voltage used for the SGM map indicated with red dots.}
    \label{fig:drawing_second_plateau}
\end{figure}

\subsubsection{Second conductance step}

In recent experiments on QPCs embedded in NS junctions it was possible to observe conductance quantization up to the second step \cite{Kjaergaard}. Here we present the change of the conductance (the difference between the conductance obtained in the presence of the SGM tip and the conductance obtained without the presence of the SGM tip) when the system is scanned with the SGM tip, choosing the value of $V_\mathrm{g}$ such that the conductance without the tip is in-between the first and second conductance plateaus (see the insets of Fig. \ref{fig:drawing_second_plateau}(a) and Fig. \ref{fig:drawing_second_plateau}(b)).

\begin{figure*}[ht!]
    \includegraphics[width=0.9\linewidth]{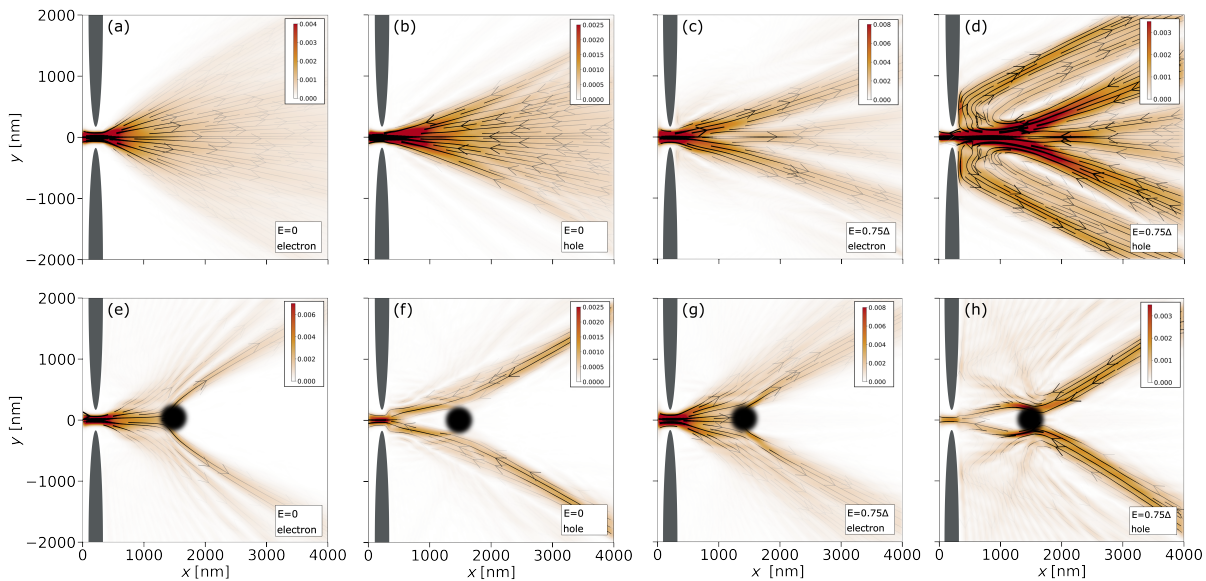}
    \caption{Probability currents of electrons (a)(c)(e)(g) and holes (b)(d)(f)(h) at second conductance slope obtained for zero energy (a)(b)(e)(f) and non-zero energy (c)(d)(g)(h). The upper row presents the results without the SGM tip, and the lower row presents the results with the SGM tip included.} 
    \label{fig:currents_second_plateau}
\end{figure*}

In Fig. \ref{fig:drawing_second_plateau}(a), the conductance change is calculated for the zero voltage bias case. In the map, a circular fringe pattern is obtained similar to the one in Fig. \ref{fig:zero energy}(a) before but now with a visible two-lobe flow from the QPC (a clear separation of the conductance fringes into two is seen between $x_\mathrm{tip} \in [500, 1000]$ nm). For the non-zero voltage bias case with $V_b = -0.75\Delta/|e|$ [Fig. \ref{fig:drawing_second_plateau}(b)], we observe a pronounced disturbance of the conductance fringes and the appearance of the resonant features. The results obtained for the QPC tuned to the second conductance step are qualitatively similar to those presented in Figs. \ref{fig:zero energy}(a) and \ref{fig:non-zero energy}(a), but it should be noted that those obtained at the first conductance step show much more pronounced features of non-retracing electron-hole trajectories due to the much simpler internal structure of the radial fringes resulting from the flow from the QPC in a single angular lobe.

Figure \ref{fig:currents_second_plateau} shows the probability current maps where $V_\mathrm{g}$ is set the same as in the conductance maps of Fig. \ref{fig:drawing_second_plateau}. For the zero-energy case, in Figs. \ref{fig:currents_second_plateau}(a) and (b), we see that the electrons are injected from QPC and the holes are retracing back the same path, although the spread of the electron flow is larger as compared to the case of Figs. \ref{fig:currents}(a, b) due to activation of the second quantization mode in the constriction. The introduction of the SGM tip, marked in the gray circle in Figs. \ref{fig:currents_second_plateau}(e)(f) deflects the paths of the electrons and holes. In the maps of Figs. \ref{fig:currents_second_plateau}(c)(d)(g)(h) we observe the same effects of the hole not retracing of the electron trajectories as in Fig. \ref{fig:currents} which leads to the disturbance of the SGM radial fringes found in Fig. \ref{fig:drawing_second_plateau}.

\section{Conclusion}
In summary, we theoretically studied the possibility of local probing of electron and hole transport in a QPC embedded in an NS junction. We proposed to use the SGM technique to observe electron and hole self-interference. We pointed out that this technique is able to unveil features of the electronic transport at non-zero bias voltage, when due to different electron and hole wavevectors the self-interference conductance oscillations exhibit beating and the Andreev-reflected hole does not retrace the electron path creating pronounced resonance patterns in the SGM probed conductance. The discussed effect occurs both at the first and also at the higher conductance step.\\

\section{Acknowledgement}
This work was supported by the National Science Center, Poland (NCN) agreement number UMO-2020/38/E/ST3/00418 and partially the program „Excellence initiative--–research university” for the AGH University of Krakow.

\bibliography{reference_revised}

\begin{thebibliography}{55}%
\makeatletter
\providecommand \@ifxundefined [1]{%
 \@ifx{#1\undefined}
}%
\providecommand \@ifnum [1]{%
 \ifnum #1\expandafter \@firstoftwo
 \else \expandafter \@secondoftwo
 \fi
}%
\providecommand \@ifx [1]{%
 \ifx #1\expandafter \@firstoftwo
 \else \expandafter \@secondoftwo
 \fi
}%
\providecommand \natexlab [1]{#1}%
\providecommand \enquote  [1]{``#1''}%
\providecommand \bibnamefont  [1]{#1}%
\providecommand \bibfnamefont [1]{#1}%
\providecommand \citenamefont [1]{#1}%
\providecommand \href@noop [0]{\@secondoftwo}%
\providecommand \href [0]{\begingroup \@sanitize@url \@href}%
\providecommand \@href[1]{\@@startlink{#1}\@@href}%
\providecommand \@@href[1]{\endgroup#1\@@endlink}%
\providecommand \@sanitize@url [0]{\catcode `\\12\catcode `\$12\catcode `\&12\catcode `\#12\catcode `\^12\catcode `\_12\catcode `\%12\relax}%
\providecommand \@@startlink[1]{}%
\providecommand \@@endlink[0]{}%
\providecommand \url  [0]{\begingroup\@sanitize@url \@url }%
\providecommand \@url [1]{\endgroup\@href {#1}{\urlprefix }}%
\providecommand \urlprefix  [0]{URL }%
\providecommand \Eprint [0]{\href }%
\providecommand \doibase [0]{https://doi.org/}%
\providecommand \selectlanguage [0]{\@gobble}%
\providecommand \bibinfo  [0]{\@secondoftwo}%
\providecommand \bibfield  [0]{\@secondoftwo}%
\providecommand \translation [1]{[#1]}%
\providecommand \BibitemOpen [0]{}%
\providecommand \bibitemStop [0]{}%
\providecommand \bibitemNoStop [0]{.\EOS\space}%
\providecommand \EOS [0]{\spacefactor3000\relax}%
\providecommand \BibitemShut  [1]{\csname bibitem#1\endcsname}%
\let\auto@bib@innerbib\@empty
\bibitem [{\citenamefont {Andreev}(1964)}]{Andreev}%
  \BibitemOpen
  \bibfield  {author} {\bibinfo {author} {\bibfnamefont {A.~F.}\ \bibnamefont {Andreev}},\ }\bibfield  {title} {\bibinfo {title} {Thermal conductivity of the intermediate state of superconductors},\ }\bibfield  {journal} {\bibinfo  {journal} {Sov. Phys. JETP.}\ }\textbf {\bibinfo {volume} {19}},\ \href {https://www.osti.gov/biblio/4071988} {} (\bibinfo {year} {1964})\BibitemShut {NoStop}%
\bibitem [{\citenamefont {Oreg}\ \emph {et~al.}(2010)\citenamefont {Oreg}, \citenamefont {Refael},\ and\ \citenamefont {von Oppen}}]{PhysRevLett.105.177002}%
  \BibitemOpen
  \bibfield  {author} {\bibinfo {author} {\bibfnamefont {Y.}~\bibnamefont {Oreg}}, \bibinfo {author} {\bibfnamefont {G.}~\bibnamefont {Refael}},\ and\ \bibinfo {author} {\bibfnamefont {F.}~\bibnamefont {von Oppen}},\ }\bibfield  {title} {\bibinfo {title} {Helical liquids and majorana bound states in quantum wires},\ }\href {https://doi.org/10.1103/PhysRevLett.105.177002} {\bibfield  {journal} {\bibinfo  {journal} {Phys. Rev. Lett.}\ }\textbf {\bibinfo {volume} {105}},\ \bibinfo {pages} {177002} (\bibinfo {year} {2010})}\BibitemShut {NoStop}%
\bibitem [{\citenamefont {Sau}\ \emph {et~al.}(2010)\citenamefont {Sau}, \citenamefont {Lutchyn}, \citenamefont {Tewari},\ and\ \citenamefont {Das~Sarma}}]{PhysRevLett.104.040502}%
  \BibitemOpen
  \bibfield  {author} {\bibinfo {author} {\bibfnamefont {J.~D.}\ \bibnamefont {Sau}}, \bibinfo {author} {\bibfnamefont {R.~M.}\ \bibnamefont {Lutchyn}}, \bibinfo {author} {\bibfnamefont {S.}~\bibnamefont {Tewari}},\ and\ \bibinfo {author} {\bibfnamefont {S.}~\bibnamefont {Das~Sarma}},\ }\bibfield  {title} {\bibinfo {title} {Generic new platform for topological quantum computation using semiconductor heterostructures},\ }\href {https://doi.org/10.1103/PhysRevLett.104.040502} {\bibfield  {journal} {\bibinfo  {journal} {Phys. Rev. Lett.}\ }\textbf {\bibinfo {volume} {104}},\ \bibinfo {pages} {040502} (\bibinfo {year} {2010})}\BibitemShut {NoStop}%
\bibitem [{\citenamefont {Mourik}\ \emph {et~al.}(2012)\citenamefont {Mourik}, \citenamefont {Zuo}, \citenamefont {Frolov}, \citenamefont {Plissard}, \citenamefont {Bakkers},\ and\ \citenamefont {Kouwenhoven}}]{Mourik2012}%
  \BibitemOpen
  \bibfield  {author} {\bibinfo {author} {\bibfnamefont {V.}~\bibnamefont {Mourik}}, \bibinfo {author} {\bibfnamefont {K.}~\bibnamefont {Zuo}}, \bibinfo {author} {\bibfnamefont {S.~M.}\ \bibnamefont {Frolov}}, \bibinfo {author} {\bibfnamefont {S.~R.}\ \bibnamefont {Plissard}}, \bibinfo {author} {\bibfnamefont {E.~P. A.~M.}\ \bibnamefont {Bakkers}},\ and\ \bibinfo {author} {\bibfnamefont {L.~P.}\ \bibnamefont {Kouwenhoven}},\ }\bibfield  {title} {\bibinfo {title} {Signatures of majorana fermions in hybrid superconductor-semiconductor nanowire devices},\ }\href {https://doi.org/10.1126/science.1222360} {\bibfield  {journal} {\bibinfo  {journal} {Science}\ }\textbf {\bibinfo {volume} {336}},\ \bibinfo {pages} {1003} (\bibinfo {year} {2012})}\BibitemShut {NoStop}%
\bibitem [{\citenamefont {Larsen}\ \emph {et~al.}(2015)\citenamefont {Larsen}, \citenamefont {Petersson}, \citenamefont {Kuemmeth}, \citenamefont {Jespersen}, \citenamefont {Krogstrup}, \citenamefont {Nyg\aa{}rd},\ and\ \citenamefont {Marcus}}]{PhysRevLett.115.127001}%
  \BibitemOpen
  \bibfield  {author} {\bibinfo {author} {\bibfnamefont {T.~W.}\ \bibnamefont {Larsen}}, \bibinfo {author} {\bibfnamefont {K.~D.}\ \bibnamefont {Petersson}}, \bibinfo {author} {\bibfnamefont {F.}~\bibnamefont {Kuemmeth}}, \bibinfo {author} {\bibfnamefont {T.~S.}\ \bibnamefont {Jespersen}}, \bibinfo {author} {\bibfnamefont {P.}~\bibnamefont {Krogstrup}}, \bibinfo {author} {\bibfnamefont {J.}~\bibnamefont {Nyg\aa{}rd}},\ and\ \bibinfo {author} {\bibfnamefont {C.~M.}\ \bibnamefont {Marcus}},\ }\bibfield  {title} {\bibinfo {title} {Semiconductor-nanowire-based superconducting qubit},\ }\href {https://doi.org/10.1103/PhysRevLett.115.127001} {\bibfield  {journal} {\bibinfo  {journal} {Phys. Rev. Lett.}\ }\textbf {\bibinfo {volume} {115}},\ \bibinfo {pages} {127001} (\bibinfo {year} {2015})}\BibitemShut {NoStop}%
\bibitem [{\citenamefont {Casparis}\ \emph {et~al.}(2018)\citenamefont {Casparis}, \citenamefont {Connolly}, \citenamefont {Kjaergaard}, \citenamefont {Pearson}, \citenamefont {Kringh{\o}j}, \citenamefont {Larsen}, \citenamefont {Kuemmeth}, \citenamefont {Wang}, \citenamefont {Thomas}, \citenamefont {Gronin}, \citenamefont {Gardner}, \citenamefont {Manfra}, \citenamefont {Marcus},\ and\ \citenamefont {Petersson}}]{Casparis2018}%
  \BibitemOpen
  \bibfield  {author} {\bibinfo {author} {\bibfnamefont {L.}~\bibnamefont {Casparis}}, \bibinfo {author} {\bibfnamefont {M.~R.}\ \bibnamefont {Connolly}}, \bibinfo {author} {\bibfnamefont {M.}~\bibnamefont {Kjaergaard}}, \bibinfo {author} {\bibfnamefont {N.~J.}\ \bibnamefont {Pearson}}, \bibinfo {author} {\bibfnamefont {A.}~\bibnamefont {Kringh{\o}j}}, \bibinfo {author} {\bibfnamefont {T.~W.}\ \bibnamefont {Larsen}}, \bibinfo {author} {\bibfnamefont {F.}~\bibnamefont {Kuemmeth}}, \bibinfo {author} {\bibfnamefont {T.}~\bibnamefont {Wang}}, \bibinfo {author} {\bibfnamefont {C.}~\bibnamefont {Thomas}}, \bibinfo {author} {\bibfnamefont {S.}~\bibnamefont {Gronin}}, \bibinfo {author} {\bibfnamefont {G.~C.}\ \bibnamefont {Gardner}}, \bibinfo {author} {\bibfnamefont {M.~J.}\ \bibnamefont {Manfra}}, \bibinfo {author} {\bibfnamefont {C.~M.}\ \bibnamefont {Marcus}},\ and\ \bibinfo {author} {\bibfnamefont {K.~D.}\ \bibnamefont {Petersson}},\ }\bibfield  {title} {\bibinfo {title} {Superconducting gatemon qubit based on a
  proximitized two-dimensional electron gas},\ }\href {https://doi.org/10.1038/s41565-018-0207-y} {\bibfield  {journal} {\bibinfo  {journal} {Nat. Nanotechnol.}\ }\textbf {\bibinfo {volume} {13}},\ \bibinfo {pages} {915} (\bibinfo {year} {2018})}\BibitemShut {NoStop}%
\bibitem [{\citenamefont {Larsen}\ \emph {et~al.}(2020)\citenamefont {Larsen}, \citenamefont {Gershenson}, \citenamefont {Casparis}, \citenamefont {Kringh\o{}j}, \citenamefont {Pearson}, \citenamefont {McNeil}, \citenamefont {Kuemmeth}, \citenamefont {Krogstrup}, \citenamefont {Petersson},\ and\ \citenamefont {Marcus}}]{PhysRevLett.125.056801}%
  \BibitemOpen
  \bibfield  {author} {\bibinfo {author} {\bibfnamefont {T.~W.}\ \bibnamefont {Larsen}}, \bibinfo {author} {\bibfnamefont {M.~E.}\ \bibnamefont {Gershenson}}, \bibinfo {author} {\bibfnamefont {L.}~\bibnamefont {Casparis}}, \bibinfo {author} {\bibfnamefont {A.}~\bibnamefont {Kringh\o{}j}}, \bibinfo {author} {\bibfnamefont {N.~J.}\ \bibnamefont {Pearson}}, \bibinfo {author} {\bibfnamefont {R.~P.~G.}\ \bibnamefont {McNeil}}, \bibinfo {author} {\bibfnamefont {F.}~\bibnamefont {Kuemmeth}}, \bibinfo {author} {\bibfnamefont {P.}~\bibnamefont {Krogstrup}}, \bibinfo {author} {\bibfnamefont {K.~D.}\ \bibnamefont {Petersson}},\ and\ \bibinfo {author} {\bibfnamefont {C.~M.}\ \bibnamefont {Marcus}},\ }\bibfield  {title} {\bibinfo {title} {Parity-protected superconductor-semiconductor qubit},\ }\href {https://doi.org/10.1103/PhysRevLett.125.056801} {\bibfield  {journal} {\bibinfo  {journal} {Phys. Rev. Lett.}\ }\textbf {\bibinfo {volume} {125}},\ \bibinfo {pages} {056801} (\bibinfo {year} {2020})}\BibitemShut {NoStop}%
\bibitem [{\citenamefont {Hays}\ \emph {et~al.}(2021)\citenamefont {Hays}, \citenamefont {Fatemi}, \citenamefont {Bouman}, \citenamefont {Cerrillo}, \citenamefont {Diamond}, \citenamefont {Serniak}, \citenamefont {Connolly}, \citenamefont {Krogstrup}, \citenamefont {Nyg{\aa}rd}, \citenamefont {Levy~Yeyati}, \citenamefont {Geresdi},\ and\ \citenamefont {Devoret}}]{Hays2021}%
  \BibitemOpen
  \bibfield  {author} {\bibinfo {author} {\bibfnamefont {M.}~\bibnamefont {Hays}}, \bibinfo {author} {\bibfnamefont {V.}~\bibnamefont {Fatemi}}, \bibinfo {author} {\bibfnamefont {D.}~\bibnamefont {Bouman}}, \bibinfo {author} {\bibfnamefont {J.}~\bibnamefont {Cerrillo}}, \bibinfo {author} {\bibfnamefont {S.}~\bibnamefont {Diamond}}, \bibinfo {author} {\bibfnamefont {K.}~\bibnamefont {Serniak}}, \bibinfo {author} {\bibfnamefont {T.}~\bibnamefont {Connolly}}, \bibinfo {author} {\bibfnamefont {P.}~\bibnamefont {Krogstrup}}, \bibinfo {author} {\bibfnamefont {J.}~\bibnamefont {Nyg{\aa}rd}}, \bibinfo {author} {\bibfnamefont {A.}~\bibnamefont {Levy~Yeyati}}, \bibinfo {author} {\bibfnamefont {A.}~\bibnamefont {Geresdi}},\ and\ \bibinfo {author} {\bibfnamefont {M.~H.}\ \bibnamefont {Devoret}},\ }\bibfield  {title} {\bibinfo {title} {Coherent manipulation of an andreev spin qubit},\ }\href {https://doi.org/10.1126/science.abf0345} {\bibfield  {journal} {\bibinfo  {journal} {Science}\ }\textbf {\bibinfo {volume} {373}},\
  \bibinfo {pages} {430} (\bibinfo {year} {2021})}\BibitemShut {NoStop}%
\bibitem [{\citenamefont {Davydova}\ \emph {et~al.}(2022)\citenamefont {Davydova}, \citenamefont {Prembabu},\ and\ \citenamefont {Fu}}]{doi:10.1126/sciadv.abo0309}%
  \BibitemOpen
  \bibfield  {author} {\bibinfo {author} {\bibfnamefont {M.}~\bibnamefont {Davydova}}, \bibinfo {author} {\bibfnamefont {S.}~\bibnamefont {Prembabu}},\ and\ \bibinfo {author} {\bibfnamefont {L.}~\bibnamefont {Fu}},\ }\bibfield  {title} {\bibinfo {title} {Universal josephson diode effect},\ }\href {https://doi.org/10.1126/sciadv.abo0309} {\bibfield  {journal} {\bibinfo  {journal} {Sci. Adv.}\ }\textbf {\bibinfo {volume} {8}},\ \bibinfo {pages} {eabo0309} (\bibinfo {year} {2022})}\BibitemShut {NoStop}%
\bibitem [{\citenamefont {Su}\ \emph {et~al.}(2017)\citenamefont {Su}, \citenamefont {Tacla}, \citenamefont {Hocevar}, \citenamefont {Car}, \citenamefont {Plissard}, \citenamefont {Bakkers}, \citenamefont {Daley}, \citenamefont {Pekker},\ and\ \citenamefont {Frolov}}]{Su2017}%
  \BibitemOpen
  \bibfield  {author} {\bibinfo {author} {\bibfnamefont {Z.}~\bibnamefont {Su}}, \bibinfo {author} {\bibfnamefont {A.~B.}\ \bibnamefont {Tacla}}, \bibinfo {author} {\bibfnamefont {M.}~\bibnamefont {Hocevar}}, \bibinfo {author} {\bibfnamefont {D.}~\bibnamefont {Car}}, \bibinfo {author} {\bibfnamefont {S.~R.}\ \bibnamefont {Plissard}}, \bibinfo {author} {\bibfnamefont {E.~P. A.~M.}\ \bibnamefont {Bakkers}}, \bibinfo {author} {\bibfnamefont {A.~J.}\ \bibnamefont {Daley}}, \bibinfo {author} {\bibfnamefont {D.}~\bibnamefont {Pekker}},\ and\ \bibinfo {author} {\bibfnamefont {S.~M.}\ \bibnamefont {Frolov}},\ }\bibfield  {title} {\bibinfo {title} {Andreev molecules in semiconductor nanowire double quantum dots},\ }\href {https://doi.org/10.1038/s41467-017-00665-7} {\bibfield  {journal} {\bibinfo  {journal} {Nat. Commun.}\ }\textbf {\bibinfo {volume} {8}},\ \bibinfo {pages} {585} (\bibinfo {year} {2017})}\BibitemShut {NoStop}%
\bibitem [{\citenamefont {Wang}\ \emph {et~al.}(2022)\citenamefont {Wang}, \citenamefont {Dvir}, \citenamefont {Mazur}, \citenamefont {Liu}, \citenamefont {van Loo}, \citenamefont {ten Haaf}, \citenamefont {Bordin}, \citenamefont {Gazibegovic}, \citenamefont {Badawy}, \citenamefont {Bakkers}, \citenamefont {Wimmer},\ and\ \citenamefont {Kouwenhoven}}]{Wang2022}%
  \BibitemOpen
  \bibfield  {author} {\bibinfo {author} {\bibfnamefont {G.}~\bibnamefont {Wang}}, \bibinfo {author} {\bibfnamefont {T.}~\bibnamefont {Dvir}}, \bibinfo {author} {\bibfnamefont {G.~P.}\ \bibnamefont {Mazur}}, \bibinfo {author} {\bibfnamefont {C.-X.}\ \bibnamefont {Liu}}, \bibinfo {author} {\bibfnamefont {N.}~\bibnamefont {van Loo}}, \bibinfo {author} {\bibfnamefont {S.~L.~D.}\ \bibnamefont {ten Haaf}}, \bibinfo {author} {\bibfnamefont {A.}~\bibnamefont {Bordin}}, \bibinfo {author} {\bibfnamefont {S.}~\bibnamefont {Gazibegovic}}, \bibinfo {author} {\bibfnamefont {G.}~\bibnamefont {Badawy}}, \bibinfo {author} {\bibfnamefont {E.~P. A.~M.}\ \bibnamefont {Bakkers}}, \bibinfo {author} {\bibfnamefont {M.}~\bibnamefont {Wimmer}},\ and\ \bibinfo {author} {\bibfnamefont {L.~P.}\ \bibnamefont {Kouwenhoven}},\ }\bibfield  {title} {\bibinfo {title} {Singlet and triplet cooper pair splitting in hybrid superconducting nanowires},\ }\href {https://doi.org/10.1038/s41586-022-05352-2} {\bibfield  {journal} {\bibinfo  {journal}
  {Nature}\ }\textbf {\bibinfo {volume} {612}},\ \bibinfo {pages} {448} (\bibinfo {year} {2022})}\BibitemShut {NoStop}%
\bibitem [{\citenamefont {Kjaergaard}\ \emph {et~al.}(2016)\citenamefont {Kjaergaard}, \citenamefont {Nichele}, \citenamefont {Suominen}, \citenamefont {Nowak}, \citenamefont {Wimmer}, \citenamefont {Akhmerov}, \citenamefont {Folk}, \citenamefont {Flensberg}, \citenamefont {Shabani}, \citenamefont {Palmstr{\o}m},\ and\ \citenamefont {Marcus}}]{Kjaergaard}%
  \BibitemOpen
  \bibfield  {author} {\bibinfo {author} {\bibfnamefont {M.}~\bibnamefont {Kjaergaard}}, \bibinfo {author} {\bibfnamefont {F.}~\bibnamefont {Nichele}}, \bibinfo {author} {\bibfnamefont {H.~J.}\ \bibnamefont {Suominen}}, \bibinfo {author} {\bibfnamefont {M.~P.}\ \bibnamefont {Nowak}}, \bibinfo {author} {\bibfnamefont {M.}~\bibnamefont {Wimmer}}, \bibinfo {author} {\bibfnamefont {A.~R.}\ \bibnamefont {Akhmerov}}, \bibinfo {author} {\bibfnamefont {J.~A.}\ \bibnamefont {Folk}}, \bibinfo {author} {\bibfnamefont {K.}~\bibnamefont {Flensberg}}, \bibinfo {author} {\bibfnamefont {J.}~\bibnamefont {Shabani}}, \bibinfo {author} {\bibfnamefont {C.~J.}\ \bibnamefont {Palmstr{\o}m}},\ and\ \bibinfo {author} {\bibfnamefont {C.~M.}\ \bibnamefont {Marcus}},\ }\bibfield  {title} {\bibinfo {title} {Quantized conductance doubling and hard gap in a two-dimensional semiconductor--superconductor heterostructure},\ }\href {https://doi.org/10.1038/ncomms12841} {\bibfield  {journal} {\bibinfo  {journal} {Nat. Commun.}\ }\textbf
  {\bibinfo {volume} {7}},\ \bibinfo {pages} {12841} (\bibinfo {year} {2016})}\BibitemShut {NoStop}%
\bibitem [{\citenamefont {Moehle}\ \emph {et~al.}(2022)\citenamefont {Moehle}, \citenamefont {Rout}, \citenamefont {Jainandunsing}, \citenamefont {Kuiri}, \citenamefont {Ke}, \citenamefont {Xiao}, \citenamefont {Thomas}, \citenamefont {Manfra}, \citenamefont {Nowak},\ and\ \citenamefont {Goswami}}]{Moehle2022}%
  \BibitemOpen
  \bibfield  {author} {\bibinfo {author} {\bibfnamefont {C.~M.}\ \bibnamefont {Moehle}}, \bibinfo {author} {\bibfnamefont {P.~K.}\ \bibnamefont {Rout}}, \bibinfo {author} {\bibfnamefont {N.~A.}\ \bibnamefont {Jainandunsing}}, \bibinfo {author} {\bibfnamefont {D.}~\bibnamefont {Kuiri}}, \bibinfo {author} {\bibfnamefont {C.~T.}\ \bibnamefont {Ke}}, \bibinfo {author} {\bibfnamefont {D.}~\bibnamefont {Xiao}}, \bibinfo {author} {\bibfnamefont {C.}~\bibnamefont {Thomas}}, \bibinfo {author} {\bibfnamefont {M.~J.}\ \bibnamefont {Manfra}}, \bibinfo {author} {\bibfnamefont {M.~P.}\ \bibnamefont {Nowak}},\ and\ \bibinfo {author} {\bibfnamefont {S.}~\bibnamefont {Goswami}},\ }\bibfield  {title} {\bibinfo {title} {Controlling andreev bound states with the magnetic vector potential},\ }\href {https://doi.org/10.1021/acs.nanolett.2c03130} {\bibfield  {journal} {\bibinfo  {journal} {Nano Lett.}\ }\textbf {\bibinfo {volume} {22}},\ \bibinfo {pages} {8601} (\bibinfo {year} {2022})}\BibitemShut {NoStop}%
\bibitem [{\citenamefont {Moehle}\ \emph {et~al.}(2021)\citenamefont {Moehle}, \citenamefont {Ke}, \citenamefont {Wang}, \citenamefont {Thomas}, \citenamefont {Xiao}, \citenamefont {Karwal}, \citenamefont {Lodari}, \citenamefont {van~de Kerkhof}, \citenamefont {Termaat}, \citenamefont {Gardner}, \citenamefont {Scappucci}, \citenamefont {Manfra},\ and\ \citenamefont {Goswami}}]{Moehle2021}%
  \BibitemOpen
  \bibfield  {author} {\bibinfo {author} {\bibfnamefont {C.~M.}\ \bibnamefont {Moehle}}, \bibinfo {author} {\bibfnamefont {C.~T.}\ \bibnamefont {Ke}}, \bibinfo {author} {\bibfnamefont {Q.}~\bibnamefont {Wang}}, \bibinfo {author} {\bibfnamefont {C.}~\bibnamefont {Thomas}}, \bibinfo {author} {\bibfnamefont {D.}~\bibnamefont {Xiao}}, \bibinfo {author} {\bibfnamefont {S.}~\bibnamefont {Karwal}}, \bibinfo {author} {\bibfnamefont {M.}~\bibnamefont {Lodari}}, \bibinfo {author} {\bibfnamefont {V.}~\bibnamefont {van~de Kerkhof}}, \bibinfo {author} {\bibfnamefont {R.}~\bibnamefont {Termaat}}, \bibinfo {author} {\bibfnamefont {G.~C.}\ \bibnamefont {Gardner}}, \bibinfo {author} {\bibfnamefont {G.}~\bibnamefont {Scappucci}}, \bibinfo {author} {\bibfnamefont {M.~J.}\ \bibnamefont {Manfra}},\ and\ \bibinfo {author} {\bibfnamefont {S.}~\bibnamefont {Goswami}},\ }\bibfield  {title} {\bibinfo {title} {Insbas two-dimensional electron gases as a platform for topological superconductivity},\ }\href
  {https://doi.org/10.1021/acs.nanolett.1c03520} {\bibfield  {journal} {\bibinfo  {journal} {Nano Lett.}\ }\textbf {\bibinfo {volume} {21}},\ \bibinfo {pages} {9990} (\bibinfo {year} {2021})}\BibitemShut {NoStop}%
\bibitem [{\citenamefont {Banerjee}\ \emph {et~al.}(2023)\citenamefont {Banerjee}, \citenamefont {Geier}, \citenamefont {Rahman}, \citenamefont {Sanchez}, \citenamefont {Thomas}, \citenamefont {Wang}, \citenamefont {Manfra}, \citenamefont {Flensberg},\ and\ \citenamefont {Marcus}}]{PhysRevLett.130.116203}%
  \BibitemOpen
  \bibfield  {author} {\bibinfo {author} {\bibfnamefont {A.}~\bibnamefont {Banerjee}}, \bibinfo {author} {\bibfnamefont {M.}~\bibnamefont {Geier}}, \bibinfo {author} {\bibfnamefont {M.~A.}\ \bibnamefont {Rahman}}, \bibinfo {author} {\bibfnamefont {D.~S.}\ \bibnamefont {Sanchez}}, \bibinfo {author} {\bibfnamefont {C.}~\bibnamefont {Thomas}}, \bibinfo {author} {\bibfnamefont {T.}~\bibnamefont {Wang}}, \bibinfo {author} {\bibfnamefont {M.~J.}\ \bibnamefont {Manfra}}, \bibinfo {author} {\bibfnamefont {K.}~\bibnamefont {Flensberg}},\ and\ \bibinfo {author} {\bibfnamefont {C.~M.}\ \bibnamefont {Marcus}},\ }\bibfield  {title} {\bibinfo {title} {Control of andreev bound states using superconducting phase texture},\ }\href {https://doi.org/10.1103/PhysRevLett.130.116203} {\bibfield  {journal} {\bibinfo  {journal} {Phys. Rev. Lett.}\ }\textbf {\bibinfo {volume} {130}},\ \bibinfo {pages} {116203} (\bibinfo {year} {2023})}\BibitemShut {NoStop}%
\bibitem [{\citenamefont {LeRoy}\ \emph {et~al.}(2002{\natexlab{a}})\citenamefont {LeRoy}, \citenamefont {Topinka}, \citenamefont {Westervelt}, \citenamefont {Maranowski},\ and\ \citenamefont {Gossard}}]{10.1063/1.1484548}%
  \BibitemOpen
  \bibfield  {author} {\bibinfo {author} {\bibfnamefont {B.~J.}\ \bibnamefont {LeRoy}}, \bibinfo {author} {\bibfnamefont {M.~A.}\ \bibnamefont {Topinka}}, \bibinfo {author} {\bibfnamefont {R.~M.}\ \bibnamefont {Westervelt}}, \bibinfo {author} {\bibfnamefont {K.~D.}\ \bibnamefont {Maranowski}},\ and\ \bibinfo {author} {\bibfnamefont {A.~C.}\ \bibnamefont {Gossard}},\ }\bibfield  {title} {\bibinfo {title} {{Imaging electron density in a two-dimensional electron gas}},\ }\href {https://doi.org/10.1063/1.1484548} {\bibfield  {journal} {\bibinfo  {journal} {Appl. Phys. Lett.}\ }\textbf {\bibinfo {volume} {80}},\ \bibinfo {pages} {4431} (\bibinfo {year} {2002}{\natexlab{a}})}\BibitemShut {NoStop}%
\bibitem [{\citenamefont {Topinka}\ \emph {et~al.}(2000)\citenamefont {Topinka}, \citenamefont {LeRoy}, \citenamefont {Shaw}, \citenamefont {Heller}, \citenamefont {Westervelt}, \citenamefont {Maranowski},\ and\ \citenamefont {Gossard}}]{Topinka2000}%
  \BibitemOpen
  \bibfield  {author} {\bibinfo {author} {\bibfnamefont {M.~A.}\ \bibnamefont {Topinka}}, \bibinfo {author} {\bibfnamefont {B.~J.}\ \bibnamefont {LeRoy}}, \bibinfo {author} {\bibfnamefont {S.~E.~J.}\ \bibnamefont {Shaw}}, \bibinfo {author} {\bibfnamefont {E.~J.}\ \bibnamefont {Heller}}, \bibinfo {author} {\bibfnamefont {R.~M.}\ \bibnamefont {Westervelt}}, \bibinfo {author} {\bibfnamefont {K.~D.}\ \bibnamefont {Maranowski}},\ and\ \bibinfo {author} {\bibfnamefont {A.~C.}\ \bibnamefont {Gossard}},\ }\bibfield  {title} {\bibinfo {title} {Imaging coherent electron flow from a quantum point contact},\ }\href {https://doi.org/10.1126/science.289.5488.2323} {\bibfield  {journal} {\bibinfo  {journal} {Science}\ }\textbf {\bibinfo {volume} {289}},\ \bibinfo {pages} {2323} (\bibinfo {year} {2000})}\BibitemShut {NoStop}%
\bibitem [{\citenamefont {Topinka}\ \emph {et~al.}(2001)\citenamefont {Topinka}, \citenamefont {LeRoy}, \citenamefont {Westervelt}, \citenamefont {Shaw}, \citenamefont {Fleischmann}, \citenamefont {Heller}, \citenamefont {Maranowski},\ and\ \citenamefont {Gossard}}]{Topinka}%
  \BibitemOpen
  \bibfield  {author} {\bibinfo {author} {\bibfnamefont {M.~A.}\ \bibnamefont {Topinka}}, \bibinfo {author} {\bibfnamefont {B.~J.}\ \bibnamefont {LeRoy}}, \bibinfo {author} {\bibfnamefont {R.~M.}\ \bibnamefont {Westervelt}}, \bibinfo {author} {\bibfnamefont {S.~E.~J.}\ \bibnamefont {Shaw}}, \bibinfo {author} {\bibfnamefont {R.}~\bibnamefont {Fleischmann}}, \bibinfo {author} {\bibfnamefont {E.~J.}\ \bibnamefont {Heller}}, \bibinfo {author} {\bibfnamefont {K.~D.}\ \bibnamefont {Maranowski}},\ and\ \bibinfo {author} {\bibfnamefont {A.~C.}\ \bibnamefont {Gossard}},\ }\bibfield  {title} {\bibinfo {title} {Coherent branched flow in a two-dimensional electron gas},\ }\href@noop {} {\bibfield  {journal} {\bibinfo  {journal} {Nature}\ }\textbf {\bibinfo {volume} {410}},\ \bibinfo {pages} {183} (\bibinfo {year} {2001})}\BibitemShut {NoStop}%
\bibitem [{\citenamefont {LeRoy}\ \emph {et~al.}(2005)\citenamefont {LeRoy}, \citenamefont {Bleszynski}, \citenamefont {Aidala}, \citenamefont {Westervelt}, \citenamefont {Kalben}, \citenamefont {Heller}, \citenamefont {Shaw}, \citenamefont {Maranowski},\ and\ \citenamefont {Gossard}}]{LeRoy}%
  \BibitemOpen
  \bibfield  {author} {\bibinfo {author} {\bibfnamefont {B.~J.}\ \bibnamefont {LeRoy}}, \bibinfo {author} {\bibfnamefont {A.~C.}\ \bibnamefont {Bleszynski}}, \bibinfo {author} {\bibfnamefont {K.~E.}\ \bibnamefont {Aidala}}, \bibinfo {author} {\bibfnamefont {R.~M.}\ \bibnamefont {Westervelt}}, \bibinfo {author} {\bibfnamefont {A.}~\bibnamefont {Kalben}}, \bibinfo {author} {\bibfnamefont {E.~J.}\ \bibnamefont {Heller}}, \bibinfo {author} {\bibfnamefont {S.~E.~J.}\ \bibnamefont {Shaw}}, \bibinfo {author} {\bibfnamefont {K.~D.}\ \bibnamefont {Maranowski}},\ and\ \bibinfo {author} {\bibfnamefont {A.~C.}\ \bibnamefont {Gossard}},\ }\bibfield  {title} {\bibinfo {title} {Imaging electron interferometer},\ }\href {https://doi.org/10.1103/PhysRevLett.94.126801} {\bibfield  {journal} {\bibinfo  {journal} {Phys. Rev. Lett.}\ }\textbf {\bibinfo {volume} {94}},\ \bibinfo {pages} {126801} (\bibinfo {year} {2005})}\BibitemShut {NoStop}%
\bibitem [{\citenamefont {Jura}\ \emph {et~al.}(2007)\citenamefont {Jura}, \citenamefont {Topinka}, \citenamefont {Urban}, \citenamefont {Yazdani}, \citenamefont {Shtrikman}, \citenamefont {Pfeiffer}, \citenamefont {West},\ and\ \citenamefont {Goldhaber-Gordon}}]{Jura_2007}%
  \BibitemOpen
  \bibfield  {author} {\bibinfo {author} {\bibfnamefont {M.~P.}\ \bibnamefont {Jura}}, \bibinfo {author} {\bibfnamefont {M.~A.}\ \bibnamefont {Topinka}}, \bibinfo {author} {\bibfnamefont {L.}~\bibnamefont {Urban}}, \bibinfo {author} {\bibfnamefont {A.}~\bibnamefont {Yazdani}}, \bibinfo {author} {\bibfnamefont {H.}~\bibnamefont {Shtrikman}}, \bibinfo {author} {\bibfnamefont {L.~N.}\ \bibnamefont {Pfeiffer}}, \bibinfo {author} {\bibfnamefont {K.~W.}\ \bibnamefont {West}},\ and\ \bibinfo {author} {\bibfnamefont {D.}~\bibnamefont {Goldhaber-Gordon}},\ }\bibfield  {title} {\bibinfo {title} {Unexpected features of branched flow through high-mobility two-dimensional electron gases},\ }\href {https://doi.org/10.1038/nphys756} {\bibfield  {journal} {\bibinfo  {journal} {Nat. Phys.}\ }\textbf {\bibinfo {volume} {3}},\ \bibinfo {pages} {841} (\bibinfo {year} {2007})}\BibitemShut {NoStop}%
\bibitem [{\citenamefont {Jura}\ \emph {et~al.}(2009)\citenamefont {Jura}, \citenamefont {Topinka}, \citenamefont {Grobis}, \citenamefont {Pfeiffer}, \citenamefont {West},\ and\ \citenamefont {Goldhaber-Gordon}}]{Electroninterferometer}%
  \BibitemOpen
  \bibfield  {author} {\bibinfo {author} {\bibfnamefont {M.~P.}\ \bibnamefont {Jura}}, \bibinfo {author} {\bibfnamefont {M.~A.}\ \bibnamefont {Topinka}}, \bibinfo {author} {\bibfnamefont {M.}~\bibnamefont {Grobis}}, \bibinfo {author} {\bibfnamefont {L.~N.}\ \bibnamefont {Pfeiffer}}, \bibinfo {author} {\bibfnamefont {K.~W.}\ \bibnamefont {West}},\ and\ \bibinfo {author} {\bibfnamefont {D.}~\bibnamefont {Goldhaber-Gordon}},\ }\bibfield  {title} {\bibinfo {title} {Electron interferometer formed with a scanning probe tip and quantum point contact},\ }\href {https://doi.org/10.1103/PhysRevB.80.041303} {\bibfield  {journal} {\bibinfo  {journal} {Phys. Rev. B}\ }\textbf {\bibinfo {volume} {80}},\ \bibinfo {pages} {041303} (\bibinfo {year} {2009})}\BibitemShut {NoStop}%
\bibitem [{\citenamefont {Jalabert}\ \emph {et~al.}(2010)\citenamefont {Jalabert}, \citenamefont {Szewc}, \citenamefont {Tomsovic},\ and\ \citenamefont {Weinmann}}]{PhysRevLett.105.166802}%
  \BibitemOpen
  \bibfield  {author} {\bibinfo {author} {\bibfnamefont {R.~A.}\ \bibnamefont {Jalabert}}, \bibinfo {author} {\bibfnamefont {W.}~\bibnamefont {Szewc}}, \bibinfo {author} {\bibfnamefont {S.}~\bibnamefont {Tomsovic}},\ and\ \bibinfo {author} {\bibfnamefont {D.}~\bibnamefont {Weinmann}},\ }\bibfield  {title} {\bibinfo {title} {What is measured in the scanning gate microscopy of a quantum point contact?},\ }\href {https://doi.org/10.1103/PhysRevLett.105.166802} {\bibfield  {journal} {\bibinfo  {journal} {Phys. Rev. Lett.}\ }\textbf {\bibinfo {volume} {105}},\ \bibinfo {pages} {166802} (\bibinfo {year} {2010})}\BibitemShut {NoStop}%
\bibitem [{\citenamefont {Aidala}\ \emph {et~al.}(2007)\citenamefont {Aidala}, \citenamefont {Parrott}, \citenamefont {Kramer}, \citenamefont {Heller}, \citenamefont {Westervelt}, \citenamefont {Hanson},\ and\ \citenamefont {Gossard}}]{Aidala2007}%
  \BibitemOpen
  \bibfield  {author} {\bibinfo {author} {\bibfnamefont {K.~E.}\ \bibnamefont {Aidala}}, \bibinfo {author} {\bibfnamefont {R.~E.}\ \bibnamefont {Parrott}}, \bibinfo {author} {\bibfnamefont {T.}~\bibnamefont {Kramer}}, \bibinfo {author} {\bibfnamefont {E.~J.}\ \bibnamefont {Heller}}, \bibinfo {author} {\bibfnamefont {R.~M.}\ \bibnamefont {Westervelt}}, \bibinfo {author} {\bibfnamefont {M.~P.}\ \bibnamefont {Hanson}},\ and\ \bibinfo {author} {\bibfnamefont {A.~C.}\ \bibnamefont {Gossard}},\ }\bibfield  {title} {\bibinfo {title} {Imaging magnetic focusing of coherent electron waves},\ }\href {https://doi.org/10.1038/nphys628} {\bibfield  {journal} {\bibinfo  {journal} {Nat. Phys.}\ }\textbf {\bibinfo {volume} {3}},\ \bibinfo {pages} {464} (\bibinfo {year} {2007})}\BibitemShut {NoStop}%
\bibitem [{\citenamefont {Bhandari}\ \emph {et~al.}(2016)\citenamefont {Bhandari}, \citenamefont {Lee}, \citenamefont {Klales}, \citenamefont {Watanabe}, \citenamefont {Taniguchi}, \citenamefont {Heller}, \citenamefont {Kim},\ and\ \citenamefont {Westervelt}}]{Bhandari2016}%
  \BibitemOpen
  \bibfield  {author} {\bibinfo {author} {\bibfnamefont {S.}~\bibnamefont {Bhandari}}, \bibinfo {author} {\bibfnamefont {G.-H.}\ \bibnamefont {Lee}}, \bibinfo {author} {\bibfnamefont {A.}~\bibnamefont {Klales}}, \bibinfo {author} {\bibfnamefont {K.}~\bibnamefont {Watanabe}}, \bibinfo {author} {\bibfnamefont {T.}~\bibnamefont {Taniguchi}}, \bibinfo {author} {\bibfnamefont {E.}~\bibnamefont {Heller}}, \bibinfo {author} {\bibfnamefont {P.}~\bibnamefont {Kim}},\ and\ \bibinfo {author} {\bibfnamefont {R.~M.}\ \bibnamefont {Westervelt}},\ }\bibfield  {title} {\bibinfo {title} {Imaging cyclotron orbits of electrons in graphene},\ }\href {https://doi.org/10.1021/acs.nanolett.5b04609} {\bibfield  {journal} {\bibinfo  {journal} {Nano Lett.}\ }\textbf {\bibinfo {volume} {16}},\ \bibinfo {pages} {1690} (\bibinfo {year} {2016})}\BibitemShut {NoStop}%
\bibitem [{\citenamefont {Bhandari}\ \emph {et~al.}(2020)\citenamefont {Bhandari}, \citenamefont {Lee}, \citenamefont {Watanabe}, \citenamefont {Taniguchi}, \citenamefont {Kim},\ and\ \citenamefont {Westervelt}}]{Bhandari2020}%
  \BibitemOpen
  \bibfield  {author} {\bibinfo {author} {\bibfnamefont {S.}~\bibnamefont {Bhandari}}, \bibinfo {author} {\bibfnamefont {G.-H.}\ \bibnamefont {Lee}}, \bibinfo {author} {\bibfnamefont {K.}~\bibnamefont {Watanabe}}, \bibinfo {author} {\bibfnamefont {T.}~\bibnamefont {Taniguchi}}, \bibinfo {author} {\bibfnamefont {P.}~\bibnamefont {Kim}},\ and\ \bibinfo {author} {\bibfnamefont {R.~M.}\ \bibnamefont {Westervelt}},\ }\bibfield  {title} {\bibinfo {title} {Imaging andreev reflection in graphene},\ }\href {https://doi.org/10.1021/acs.nanolett.0c00903} {\bibfield  {journal} {\bibinfo  {journal} {Nano Lett.}\ }\textbf {\bibinfo {volume} {20}},\ \bibinfo {pages} {4890} (\bibinfo {year} {2020})}\BibitemShut {NoStop}%
\bibitem [{\citenamefont {Kolasi\ifmmode~\acute{n}\else \'{n}\fi{}ski}\ \emph {et~al.}(2014)\citenamefont {Kolasi\ifmmode~\acute{n}\else \'{n}\fi{}ski}, \citenamefont {Szafran},\ and\ \citenamefont {Nowak}}]{PhysRevB.90.165303}%
  \BibitemOpen
  \bibfield  {author} {\bibinfo {author} {\bibfnamefont {K.}~\bibnamefont {Kolasi\ifmmode~\acute{n}\else \'{n}\fi{}ski}}, \bibinfo {author} {\bibfnamefont {B.}~\bibnamefont {Szafran}},\ and\ \bibinfo {author} {\bibfnamefont {M.~P.}\ \bibnamefont {Nowak}},\ }\bibfield  {title} {\bibinfo {title} {Imaging of double slit interference by scanning gate microscopy},\ }\href {https://doi.org/10.1103/PhysRevB.90.165303} {\bibfield  {journal} {\bibinfo  {journal} {Phys. Rev. B}\ }\textbf {\bibinfo {volume} {90}},\ \bibinfo {pages} {165303} (\bibinfo {year} {2014})}\BibitemShut {NoStop}%
\bibitem [{\citenamefont {Nowak}\ \emph {et~al.}(2014)\citenamefont {Nowak}, \citenamefont {Kolasi\ifmmode~\acute{n}\else \'{n}\fi{}ski},\ and\ \citenamefont {Szafran}}]{PhysRevB.90.035301}%
  \BibitemOpen
  \bibfield  {author} {\bibinfo {author} {\bibfnamefont {M.~P.}\ \bibnamefont {Nowak}}, \bibinfo {author} {\bibfnamefont {K.}~\bibnamefont {Kolasi\ifmmode~\acute{n}\else \'{n}\fi{}ski}},\ and\ \bibinfo {author} {\bibfnamefont {B.}~\bibnamefont {Szafran}},\ }\bibfield  {title} {\bibinfo {title} {Signatures of spin-orbit coupling in scanning gate conductance images of electron flow from quantum point contacts},\ }\href {https://doi.org/10.1103/PhysRevB.90.035301} {\bibfield  {journal} {\bibinfo  {journal} {Phys. Rev. B}\ }\textbf {\bibinfo {volume} {90}},\ \bibinfo {pages} {035301} (\bibinfo {year} {2014})}\BibitemShut {NoStop}%
\bibitem [{\citenamefont {Kolasi\ifmmode~\acute{n}\else \'{n}\fi{}ski}\ \emph {et~al.}(2016)\citenamefont {Kolasi\ifmmode~\acute{n}\else \'{n}\fi{}ski}, \citenamefont {Szafran}, \citenamefont {Brun},\ and\ \citenamefont {Sellier}}]{PhysRevB.94.075301}%
  \BibitemOpen
  \bibfield  {author} {\bibinfo {author} {\bibfnamefont {K.}~\bibnamefont {Kolasi\ifmmode~\acute{n}\else \'{n}\fi{}ski}}, \bibinfo {author} {\bibfnamefont {B.}~\bibnamefont {Szafran}}, \bibinfo {author} {\bibfnamefont {B.}~\bibnamefont {Brun}},\ and\ \bibinfo {author} {\bibfnamefont {H.}~\bibnamefont {Sellier}},\ }\bibfield  {title} {\bibinfo {title} {Interference features in scanning gate conductance maps of quantum point contacts with disorder},\ }\href {https://doi.org/10.1103/PhysRevB.94.075301} {\bibfield  {journal} {\bibinfo  {journal} {Phys. Rev. B}\ }\textbf {\bibinfo {volume} {94}},\ \bibinfo {pages} {075301} (\bibinfo {year} {2016})}\BibitemShut {NoStop}%
\bibitem [{\citenamefont {Mre\ifmmode \acute{n}\else \'{n}\fi{}ca-Kolasi\ifmmode~\acute{n}\else \'{n}\fi{}ska}\ \emph {et~al.}(2018)\citenamefont {Mre\ifmmode \acute{n}\else \'{n}\fi{}ca-Kolasi\ifmmode~\acute{n}\else \'{n}\fi{}ska}, \citenamefont {Kolasi\ifmmode~\acute{n}\else \'{n}\fi{}ski},\ and\ \citenamefont {Szafran}}]{PhysRevB.98.115309}%
  \BibitemOpen
  \bibfield  {author} {\bibinfo {author} {\bibfnamefont {A.}~\bibnamefont {Mre\ifmmode \acute{n}\else \'{n}\fi{}ca-Kolasi\ifmmode~\acute{n}\else \'{n}\fi{}ska}}, \bibinfo {author} {\bibfnamefont {K.}~\bibnamefont {Kolasi\ifmmode~\acute{n}\else \'{n}\fi{}ski}},\ and\ \bibinfo {author} {\bibfnamefont {B.}~\bibnamefont {Szafran}},\ }\bibfield  {title} {\bibinfo {title} {Imaging spin-resolved cyclotron trajectories in the insb two-dimensional electron gas},\ }\href {https://doi.org/10.1103/PhysRevB.98.115309} {\bibfield  {journal} {\bibinfo  {journal} {Phys. Rev. B}\ }\textbf {\bibinfo {volume} {98}},\ \bibinfo {pages} {115309} (\bibinfo {year} {2018})}\BibitemShut {NoStop}%
\bibitem [{\citenamefont {Mre\ifmmode \acute{n}\else \'{n}\fi{}ca-Kolasi\ifmmode~\acute{n}\else \'{n}\fi{}ska}\ and\ \citenamefont {Szafran}(2017)}]{PhysRevB.96.165310}%
  \BibitemOpen
  \bibfield  {author} {\bibinfo {author} {\bibfnamefont {A.}~\bibnamefont {Mre\ifmmode \acute{n}\else \'{n}\fi{}ca-Kolasi\ifmmode~\acute{n}\else \'{n}\fi{}ska}}\ and\ \bibinfo {author} {\bibfnamefont {B.}~\bibnamefont {Szafran}},\ }\bibfield  {title} {\bibinfo {title} {Imaging backscattering in graphene quantum point contacts},\ }\href {https://doi.org/10.1103/PhysRevB.96.165310} {\bibfield  {journal} {\bibinfo  {journal} {Phys. Rev. B}\ }\textbf {\bibinfo {volume} {96}},\ \bibinfo {pages} {165310} (\bibinfo {year} {2017})}\BibitemShut {NoStop}%
\bibitem [{\citenamefont {Kolasi\ifmmode~\acute{n}\else \'{n}\fi{}ski}\ \emph {et~al.}(2017)\citenamefont {Kolasi\ifmmode~\acute{n}\else \'{n}\fi{}ski}, \citenamefont {Mre\ifmmode \acute{n}\else \'{n}\fi{}ca-Kolasi\ifmmode~\acute{n}\else \'{n}\fi{}ska},\ and\ \citenamefont {Szafran}}]{PhysRevB.95.045304}%
  \BibitemOpen
  \bibfield  {author} {\bibinfo {author} {\bibfnamefont {K.}~\bibnamefont {Kolasi\ifmmode~\acute{n}\else \'{n}\fi{}ski}}, \bibinfo {author} {\bibfnamefont {A.}~\bibnamefont {Mre\ifmmode \acute{n}\else \'{n}\fi{}ca-Kolasi\ifmmode~\acute{n}\else \'{n}\fi{}ska}},\ and\ \bibinfo {author} {\bibfnamefont {B.}~\bibnamefont {Szafran}},\ }\bibfield  {title} {\bibinfo {title} {Imaging snake orbits at graphene $n\ensuremath{-}p$ junctions},\ }\href {https://doi.org/10.1103/PhysRevB.95.045304} {\bibfield  {journal} {\bibinfo  {journal} {Phys. Rev. B}\ }\textbf {\bibinfo {volume} {95}},\ \bibinfo {pages} {045304} (\bibinfo {year} {2017})}\BibitemShut {NoStop}%
\bibitem [{\citenamefont {Mreńca}\ \emph {et~al.}(2015)\citenamefont {Mreńca}, \citenamefont {Kolasiński},\ and\ \citenamefont {Szafran}}]{Mrenca_2015}%
  \BibitemOpen
  \bibfield  {author} {\bibinfo {author} {\bibfnamefont {A.}~\bibnamefont {Mreńca}}, \bibinfo {author} {\bibfnamefont {K.}~\bibnamefont {Kolasiński}},\ and\ \bibinfo {author} {\bibfnamefont {B.}~\bibnamefont {Szafran}},\ }\bibfield  {title} {\bibinfo {title} {Conductance response of graphene nanoribbons and quantum point contacts in scanning gate measurements},\ }\href {https://doi.org/10.1088/0268-1242/30/8/085003} {\bibfield  {journal} {\bibinfo  {journal} {Semicond Sci Technol}\ }\textbf {\bibinfo {volume} {30}},\ \bibinfo {pages} {085003} (\bibinfo {year} {2015})}\BibitemShut {NoStop}%
\bibitem [{\citenamefont {Prokop}\ \emph {et~al.}(2020)\citenamefont {Prokop}, \citenamefont {Gut},\ and\ \citenamefont {Nowak}}]{Prokop_2020}%
  \BibitemOpen
  \bibfield  {author} {\bibinfo {author} {\bibfnamefont {M.}~\bibnamefont {Prokop}}, \bibinfo {author} {\bibfnamefont {D.}~\bibnamefont {Gut}},\ and\ \bibinfo {author} {\bibfnamefont {M.~P.}\ \bibnamefont {Nowak}},\ }\bibfield  {title} {\bibinfo {title} {Scanning gate microscopy mapping of edge current and branched electron flow in a transition metal dichalcogenide nanoribbon and quantum point contact},\ }\href {https://doi.org/10.1088/1361-648X/ab6f83} {\bibfield  {journal} {\bibinfo  {journal} {J. Condens. Matter Phys.}\ }\textbf {\bibinfo {volume} {32}},\ \bibinfo {pages} {205302} (\bibinfo {year} {2020})}\BibitemShut {NoStop}%
\bibitem [{\citenamefont {Beenakker}(1992)}]{Beenakker}%
  \BibitemOpen
  \bibfield  {author} {\bibinfo {author} {\bibfnamefont {C.~W.~J.}\ \bibnamefont {Beenakker}},\ }\bibfield  {title} {\bibinfo {title} {Quantum transport in semiconductor-superconductor microjunctions},\ }\href {https://doi.org/10.1103/PhysRevB.46.12841} {\bibfield  {journal} {\bibinfo  {journal} {Phys. Rev. B}\ }\textbf {\bibinfo {volume} {46}},\ \bibinfo {pages} {12841} (\bibinfo {year} {1992})}\BibitemShut {NoStop}%
\bibitem [{\citenamefont {Zhang}\ \emph {et~al.}(2017)\citenamefont {Zhang}, \citenamefont {G{\"u}l}, \citenamefont {Conesa-Boj}, \citenamefont {Nowak}, \citenamefont {Wimmer}, \citenamefont {Zuo}, \citenamefont {Mourik}, \citenamefont {de~Vries}, \citenamefont {van Veen}, \citenamefont {de~Moor}, \citenamefont {Bommer}, \citenamefont {van Woerkom}, \citenamefont {Car}, \citenamefont {Plissard}, \citenamefont {Bakkers}, \citenamefont {Quintero-P{\'e}rez}, \citenamefont {Cassidy}, \citenamefont {Koelling}, \citenamefont {Goswami}, \citenamefont {Watanabe}, \citenamefont {Taniguchi},\ and\ \citenamefont {Kouwenhoven}}]{Zhang2017}%
  \BibitemOpen
  \bibfield  {author} {\bibinfo {author} {\bibfnamefont {H.}~\bibnamefont {Zhang}}, \bibinfo {author} {\bibfnamefont {{\"O}.}~\bibnamefont {G{\"u}l}}, \bibinfo {author} {\bibfnamefont {S.}~\bibnamefont {Conesa-Boj}}, \bibinfo {author} {\bibfnamefont {M.~P.}\ \bibnamefont {Nowak}}, \bibinfo {author} {\bibfnamefont {M.}~\bibnamefont {Wimmer}}, \bibinfo {author} {\bibfnamefont {K.}~\bibnamefont {Zuo}}, \bibinfo {author} {\bibfnamefont {V.}~\bibnamefont {Mourik}}, \bibinfo {author} {\bibfnamefont {F.~K.}\ \bibnamefont {de~Vries}}, \bibinfo {author} {\bibfnamefont {J.}~\bibnamefont {van Veen}}, \bibinfo {author} {\bibfnamefont {M.~W.~A.}\ \bibnamefont {de~Moor}}, \bibinfo {author} {\bibfnamefont {J.~D.~S.}\ \bibnamefont {Bommer}}, \bibinfo {author} {\bibfnamefont {D.~J.}\ \bibnamefont {van Woerkom}}, \bibinfo {author} {\bibfnamefont {D.}~\bibnamefont {Car}}, \bibinfo {author} {\bibfnamefont {S.~R.}\ \bibnamefont {Plissard}}, \bibinfo {author} {\bibfnamefont {E.~P.}\ \bibnamefont {Bakkers}}, \bibinfo {author}
  {\bibfnamefont {M.}~\bibnamefont {Quintero-P{\'e}rez}}, \bibinfo {author} {\bibfnamefont {M.~C.}\ \bibnamefont {Cassidy}}, \bibinfo {author} {\bibfnamefont {S.}~\bibnamefont {Koelling}}, \bibinfo {author} {\bibfnamefont {S.}~\bibnamefont {Goswami}}, \bibinfo {author} {\bibfnamefont {K.}~\bibnamefont {Watanabe}}, \bibinfo {author} {\bibfnamefont {T.}~\bibnamefont {Taniguchi}},\ and\ \bibinfo {author} {\bibfnamefont {L.~P.}\ \bibnamefont {Kouwenhoven}},\ }\bibfield  {title} {\bibinfo {title} {Ballistic superconductivity in semiconductor nanowires},\ }\href {https://doi.org/10.1038/ncomms16025} {\bibfield  {journal} {\bibinfo  {journal} {Nat. Commun.}\ }\textbf {\bibinfo {volume} {8}},\ \bibinfo {pages} {16025} (\bibinfo {year} {2017})}\BibitemShut {NoStop}%
\bibitem [{\citenamefont {Libisch}\ \emph {et~al.}(2006)\citenamefont {Libisch}, \citenamefont {Rotter},\ and\ \citenamefont {Burgd\"orfer}}]{PhysRevB.73.045324}%
  \BibitemOpen
  \bibfield  {author} {\bibinfo {author} {\bibfnamefont {F.}~\bibnamefont {Libisch}}, \bibinfo {author} {\bibfnamefont {S.}~\bibnamefont {Rotter}},\ and\ \bibinfo {author} {\bibfnamefont {J.}~\bibnamefont {Burgd\"orfer}},\ }\bibfield  {title} {\bibinfo {title} {Non-retracing orbits in andreev billiards},\ }\href {https://doi.org/10.1103/PhysRevB.73.045324} {\bibfield  {journal} {\bibinfo  {journal} {Phys. Rev. B}\ }\textbf {\bibinfo {volume} {73}},\ \bibinfo {pages} {045324} (\bibinfo {year} {2006})}\BibitemShut {NoStop}%
\bibitem [{\citenamefont {Korm\'anyos}\ \emph {et~al.}(2006)\citenamefont {Korm\'anyos}, \citenamefont {Kaufmann}, \citenamefont {Cserti},\ and\ \citenamefont {Lambert}}]{PhysRevLett.96.237002}%
  \BibitemOpen
  \bibfield  {author} {\bibinfo {author} {\bibfnamefont {A.}~\bibnamefont {Korm\'anyos}}, \bibinfo {author} {\bibfnamefont {Z.}~\bibnamefont {Kaufmann}}, \bibinfo {author} {\bibfnamefont {J.}~\bibnamefont {Cserti}},\ and\ \bibinfo {author} {\bibfnamefont {C.~J.}\ \bibnamefont {Lambert}},\ }\bibfield  {title} {\bibinfo {title} {Quantum-classical correspondence in the wave functions of andreev billiards},\ }\href {https://doi.org/10.1103/PhysRevLett.96.237002} {\bibfield  {journal} {\bibinfo  {journal} {Phys. Rev. Lett.}\ }\textbf {\bibinfo {volume} {96}},\ \bibinfo {pages} {237002} (\bibinfo {year} {2006})}\BibitemShut {NoStop}%
\bibitem [{\citenamefont {Libisch}\ \emph {et~al.}(2007)\citenamefont {Libisch}, \citenamefont {Rotter},\ and\ \citenamefont {Burgd{\"o}rfer}}]{Libisch2007}%
  \BibitemOpen
  \bibfield  {author} {\bibinfo {author} {\bibfnamefont {F.}~\bibnamefont {Libisch}}, \bibinfo {author} {\bibfnamefont {S.}~\bibnamefont {Rotter}},\ and\ \bibinfo {author} {\bibfnamefont {J.}~\bibnamefont {Burgd{\"o}rfer}},\ }\bibfield  {title} {\bibinfo {title} {Chladni figures in andreev billiards},\ }\href {https://doi.org/10.1140/epjst/e2007-00160-5} {\bibfield  {journal} {\bibinfo  {journal} {Eur. Phys. J.: Spec. Top.}\ }\textbf {\bibinfo {volume} {145}},\ \bibinfo {pages} {245} (\bibinfo {year} {2007})}\BibitemShut {NoStop}%
\bibitem [{\citenamefont {Kaperek}\ \emph {et~al.}(2022)\citenamefont {Kaperek}, \citenamefont {Heun}, \citenamefont {Carrega}, \citenamefont {W\'ojcik},\ and\ \citenamefont {Nowak}}]{PhysRevB.106.035432}%
  \BibitemOpen
  \bibfield  {author} {\bibinfo {author} {\bibfnamefont {K.}~\bibnamefont {Kaperek}}, \bibinfo {author} {\bibfnamefont {S.}~\bibnamefont {Heun}}, \bibinfo {author} {\bibfnamefont {M.}~\bibnamefont {Carrega}}, \bibinfo {author} {\bibfnamefont {P.}~\bibnamefont {W\'ojcik}},\ and\ \bibinfo {author} {\bibfnamefont {M.~P.}\ \bibnamefont {Nowak}},\ }\bibfield  {title} {\bibinfo {title} {Theory of scanning gate microscopy imaging of the supercurrent distribution in a planar josephson junction},\ }\href {https://doi.org/10.1103/PhysRevB.106.035432} {\bibfield  {journal} {\bibinfo  {journal} {Phys. Rev. B}\ }\textbf {\bibinfo {volume} {106}},\ \bibinfo {pages} {035432} (\bibinfo {year} {2022})}\BibitemShut {NoStop}%
\bibitem [{\citenamefont {Davies}\ \emph {et~al.}(1995)\citenamefont {Davies}, \citenamefont {Larkin},\ and\ \citenamefont {Sukhorukov}}]{Davies}%
  \BibitemOpen
  \bibfield  {author} {\bibinfo {author} {\bibfnamefont {J.~H.}\ \bibnamefont {Davies}}, \bibinfo {author} {\bibfnamefont {I.~A.}\ \bibnamefont {Larkin}},\ and\ \bibinfo {author} {\bibfnamefont {E.~V.}\ \bibnamefont {Sukhorukov}},\ }\bibfield  {title} {\bibinfo {title} {Modeling the patterned two‐dimensional electron gas: Electrostatics},\ }\href {https://doi.org/10.1063/1.359446} {\bibfield  {journal} {\bibinfo  {journal} {J. Appl. Phys.}\ }\textbf {\bibinfo {volume} {77}},\ \bibinfo {pages} {4504} (\bibinfo {year} {1995})}\BibitemShut {NoStop}%
\bibitem [{\citenamefont {Aoki}\ \emph {et~al.}(2005)\citenamefont {Aoki}, \citenamefont {Da~Cunha}, \citenamefont {Akis}, \citenamefont {Ferry},\ and\ \citenamefont {Ochiai}}]{Aoki}%
  \BibitemOpen
  \bibfield  {author} {\bibinfo {author} {\bibfnamefont {N.}~\bibnamefont {Aoki}}, \bibinfo {author} {\bibfnamefont {C.~R.}\ \bibnamefont {Da~Cunha}}, \bibinfo {author} {\bibfnamefont {R.}~\bibnamefont {Akis}}, \bibinfo {author} {\bibfnamefont {D.~K.}\ \bibnamefont {Ferry}},\ and\ \bibinfo {author} {\bibfnamefont {Y.}~\bibnamefont {Ochiai}},\ }\bibfield  {title} {\bibinfo {title} {Scanning gate microscopy investigations on an ingaas quantum point contact},\ }\href {https://doi.org/10.1063/1.2136408} {\bibfield  {journal} {\bibinfo  {journal} {Appl. Phys. Lett.}\ }\textbf {\bibinfo {volume} {87}},\ \bibinfo {pages} {223501} (\bibinfo {year} {2005})}\BibitemShut {NoStop}%
\bibitem [{\citenamefont {Iagallo}\ \emph {et~al.}(2015)\citenamefont {Iagallo}, \citenamefont {Paradiso}, \citenamefont {Roddaro}, \citenamefont {Reichl}, \citenamefont {Wegscheider}, \citenamefont {Biasiol}, \citenamefont {Sorba}, \citenamefont {Beltram},\ and\ \citenamefont {Heun}}]{Iagallo}%
  \BibitemOpen
  \bibfield  {author} {\bibinfo {author} {\bibfnamefont {A.}~\bibnamefont {Iagallo}}, \bibinfo {author} {\bibfnamefont {N.}~\bibnamefont {Paradiso}}, \bibinfo {author} {\bibfnamefont {S.}~\bibnamefont {Roddaro}}, \bibinfo {author} {\bibfnamefont {C.}~\bibnamefont {Reichl}}, \bibinfo {author} {\bibfnamefont {W.}~\bibnamefont {Wegscheider}}, \bibinfo {author} {\bibfnamefont {G.}~\bibnamefont {Biasiol}}, \bibinfo {author} {\bibfnamefont {L.}~\bibnamefont {Sorba}}, \bibinfo {author} {\bibfnamefont {F.}~\bibnamefont {Beltram}},\ and\ \bibinfo {author} {\bibfnamefont {S.}~\bibnamefont {Heun}},\ }\bibfield  {title} {\bibinfo {title} {Scanning gate imaging of quantum point contacts and the origin of the 0.7 anomaly},\ }\href {https://doi.org/10.1007/s12274-014-0576-y} {\bibfield  {journal} {\bibinfo  {journal} {Nano Res.}\ }\textbf {\bibinfo {volume} {8}},\ \bibinfo {pages} {948} (\bibinfo {year} {2015})}\BibitemShut {NoStop}%
\bibitem [{\citenamefont {Szafran}(2011)}]{Szafran}%
  \BibitemOpen
  \bibfield  {author} {\bibinfo {author} {\bibfnamefont {B.}~\bibnamefont {Szafran}},\ }\bibfield  {title} {\bibinfo {title} {Scanning gate microscopy simulations for quantum rings: Effective potential of the tip and conductance maps},\ }\href {https://doi.org/10.1103/PhysRevB.84.075336} {\bibfield  {journal} {\bibinfo  {journal} {Phys. Rev. B}\ }\textbf {\bibinfo {volume} {84}},\ \bibinfo {pages} {075336} (\bibinfo {year} {2011})}\BibitemShut {NoStop}%
\bibitem [{lin()}]{linear_note}%
  \BibitemOpen
  \href@noop {} {}\bibinfo {note} {The linear response approach, is capable of reproducing the characteristic condauctance features obtained experimentally at the finite bias \cite{Electroninterferometer, leroy2002imaging}.}\BibitemShut {Stop}%
\bibitem [{\citenamefont {Rosdahl}\ \emph {et~al.}(2018)\citenamefont {Rosdahl}, \citenamefont {Vuik}, \citenamefont {Kjaergaard},\ and\ \citenamefont {Akhmerov}}]{PhysRevB.97.045421}%
  \BibitemOpen
  \bibfield  {author} {\bibinfo {author} {\bibfnamefont {T.~O.}\ \bibnamefont {Rosdahl}}, \bibinfo {author} {\bibfnamefont {A.}~\bibnamefont {Vuik}}, \bibinfo {author} {\bibfnamefont {M.}~\bibnamefont {Kjaergaard}},\ and\ \bibinfo {author} {\bibfnamefont {A.~R.}\ \bibnamefont {Akhmerov}},\ }\bibfield  {title} {\bibinfo {title} {Andreev rectifier: A nonlocal conductance signature of topological phase transitions},\ }\href {https://doi.org/10.1103/PhysRevB.97.045421} {\bibfield  {journal} {\bibinfo  {journal} {Phys. Rev. B}\ }\textbf {\bibinfo {volume} {97}},\ \bibinfo {pages} {045421} (\bibinfo {year} {2018})}\BibitemShut {NoStop}%
\bibitem [{\citenamefont {Salimian}\ \emph {et~al.}(2021)\citenamefont {Salimian}, \citenamefont {Carrega}, \citenamefont {Verma}, \citenamefont {Zannier}, \citenamefont {Nowak}, \citenamefont {Beltram}, \citenamefont {Sorba},\ and\ \citenamefont {Heun}}]{Salimian}%
  \BibitemOpen
  \bibfield  {author} {\bibinfo {author} {\bibfnamefont {S.}~\bibnamefont {Salimian}}, \bibinfo {author} {\bibfnamefont {M.}~\bibnamefont {Carrega}}, \bibinfo {author} {\bibfnamefont {I.}~\bibnamefont {Verma}}, \bibinfo {author} {\bibfnamefont {V.}~\bibnamefont {Zannier}}, \bibinfo {author} {\bibfnamefont {M.~P.}\ \bibnamefont {Nowak}}, \bibinfo {author} {\bibfnamefont {F.}~\bibnamefont {Beltram}}, \bibinfo {author} {\bibfnamefont {L.}~\bibnamefont {Sorba}},\ and\ \bibinfo {author} {\bibfnamefont {S.}~\bibnamefont {Heun}},\ }\bibfield  {title} {\bibinfo {title} {{Gate-controlled supercurrent in ballistic InSb nanoflag Josephson junctions}},\ }\href {https://doi.org/10.1063/5.0071218} {\bibfield  {journal} {\bibinfo  {journal} {Appl. Phys. Lett.}\ }\textbf {\bibinfo {volume} {119}},\ \bibinfo {pages} {214004} (\bibinfo {year} {2021})}\BibitemShut {NoStop}%
\bibitem [{\citenamefont {Zhi}\ \emph {et~al.}(2019{\natexlab{a}})\citenamefont {Zhi}, \citenamefont {Kang}, \citenamefont {Li}, \citenamefont {Fan}, \citenamefont {Su}, \citenamefont {Pan}, \citenamefont {Zhao}, \citenamefont {Zhao},\ and\ \citenamefont {Xu}}]{Zhi}%
  \BibitemOpen
  \bibfield  {author} {\bibinfo {author} {\bibfnamefont {J.}~\bibnamefont {Zhi}}, \bibinfo {author} {\bibfnamefont {N.}~\bibnamefont {Kang}}, \bibinfo {author} {\bibfnamefont {S.}~\bibnamefont {Li}}, \bibinfo {author} {\bibfnamefont {D.}~\bibnamefont {Fan}}, \bibinfo {author} {\bibfnamefont {F.}~\bibnamefont {Su}}, \bibinfo {author} {\bibfnamefont {D.}~\bibnamefont {Pan}}, \bibinfo {author} {\bibfnamefont {S.}~\bibnamefont {Zhao}}, \bibinfo {author} {\bibfnamefont {J.}~\bibnamefont {Zhao}},\ and\ \bibinfo {author} {\bibfnamefont {H.}~\bibnamefont {Xu}},\ }\bibfield  {title} {\bibinfo {title} {Supercurrent and multiple andreev reflections in insb nanosheet sns junctions},\ }\href {https://doi.org/https://doi.org/10.1002/pssb.201800538} {\bibfield  {journal} {\bibinfo  {journal} {Phys. Status Solidi B}\ }\textbf {\bibinfo {volume} {256}},\ \bibinfo {pages} {1800538} (\bibinfo {year} {2019}{\natexlab{a}})}\BibitemShut {NoStop}%
\bibitem [{\citenamefont {Zhi}\ \emph {et~al.}(2019{\natexlab{b}})\citenamefont {Zhi}, \citenamefont {Kang}, \citenamefont {Su}, \citenamefont {Fan}, \citenamefont {Li}, \citenamefont {Pan}, \citenamefont {Zhao}, \citenamefont {Zhao},\ and\ \citenamefont {Xu}}]{PhysRevB.99.245302}%
  \BibitemOpen
  \bibfield  {author} {\bibinfo {author} {\bibfnamefont {J.}~\bibnamefont {Zhi}}, \bibinfo {author} {\bibfnamefont {N.}~\bibnamefont {Kang}}, \bibinfo {author} {\bibfnamefont {F.}~\bibnamefont {Su}}, \bibinfo {author} {\bibfnamefont {D.}~\bibnamefont {Fan}}, \bibinfo {author} {\bibfnamefont {S.}~\bibnamefont {Li}}, \bibinfo {author} {\bibfnamefont {D.}~\bibnamefont {Pan}}, \bibinfo {author} {\bibfnamefont {S.~P.}\ \bibnamefont {Zhao}}, \bibinfo {author} {\bibfnamefont {J.}~\bibnamefont {Zhao}},\ and\ \bibinfo {author} {\bibfnamefont {H.~Q.}\ \bibnamefont {Xu}},\ }\bibfield  {title} {\bibinfo {title} {Coexistence of induced superconductivity and quantum hall states in insb nanosheets},\ }\href {https://doi.org/10.1103/PhysRevB.99.245302} {\bibfield  {journal} {\bibinfo  {journal} {Phys. Rev. B}\ }\textbf {\bibinfo {volume} {99}},\ \bibinfo {pages} {245302} (\bibinfo {year} {2019}{\natexlab{b}})}\BibitemShut {NoStop}%
\bibitem [{\citenamefont {Satchell}\ \emph {et~al.}(2023)\citenamefont {Satchell}, \citenamefont {Shepley}, \citenamefont {Rosamond},\ and\ \citenamefont {Burnell}}]{Satchell}%
  \BibitemOpen
  \bibfield  {author} {\bibinfo {author} {\bibfnamefont {N.}~\bibnamefont {Satchell}}, \bibinfo {author} {\bibfnamefont {P.}~\bibnamefont {Shepley}}, \bibinfo {author} {\bibfnamefont {M.}~\bibnamefont {Rosamond}},\ and\ \bibinfo {author} {\bibfnamefont {G.}~\bibnamefont {Burnell}},\ }\bibfield  {title} {\bibinfo {title} {{Supercurrent diode effect in thin film Nb tracks}},\ }\href {https://doi.org/10.1063/5.0141576} {\bibfield  {journal} {\bibinfo  {journal} {J. Appl. Phys.}\ }\textbf {\bibinfo {volume} {133}},\ \bibinfo {pages} {203901} (\bibinfo {year} {2023})}\BibitemShut {NoStop}%
\bibitem [{\citenamefont {Groth}\ \emph {et~al.}(2014)\citenamefont {Groth}, \citenamefont {Wimmer}, \citenamefont {Akhmerov},\ and\ \citenamefont {Waintal}}]{Christoph}%
  \BibitemOpen
  \bibfield  {author} {\bibinfo {author} {\bibfnamefont {C.~W.}\ \bibnamefont {Groth}}, \bibinfo {author} {\bibfnamefont {M.}~\bibnamefont {Wimmer}}, \bibinfo {author} {\bibfnamefont {A.~R.}\ \bibnamefont {Akhmerov}},\ and\ \bibinfo {author} {\bibfnamefont {X.}~\bibnamefont {Waintal}},\ }\bibfield  {title} {\bibinfo {title} {Kwant: a software package for quantum transport},\ }\href {https://doi.org/10.1088/1367-2630/16/6/063065} {\bibfield  {journal} {\bibinfo  {journal} {New J. Phys.}\ }\textbf {\bibinfo {volume} {16}},\ \bibinfo {pages} {063065} (\bibinfo {year} {2014})}\BibitemShut {NoStop}%
\bibitem [{\citenamefont {Nijholt}\ \emph {et~al.}(2019)\citenamefont {Nijholt}, \citenamefont {Weston}, \citenamefont {Hoofwijk},\ and\ \citenamefont {Akhmerov}}]{Nijholt2019}%
  \BibitemOpen
  \bibfield  {author} {\bibinfo {author} {\bibfnamefont {B.}~\bibnamefont {Nijholt}}, \bibinfo {author} {\bibfnamefont {J.}~\bibnamefont {Weston}}, \bibinfo {author} {\bibfnamefont {J.}~\bibnamefont {Hoofwijk}},\ and\ \bibinfo {author} {\bibfnamefont {A.}~\bibnamefont {Akhmerov}},\ }\href {https://doi.org/10.5281/zenodo.1182437} {\bibinfo {title} {\textit{Adaptive}: parallel active learning of mathematical functions}} (\bibinfo {year} {2019})\BibitemShut {NoStop}%
\bibitem [{\citenamefont {Maji}\ and\ \citenamefont {Nowak}(2023)}]{zenodo_repository}%
  \BibitemOpen
  \bibfield  {author} {\bibinfo {author} {\bibfnamefont {S.}~\bibnamefont {Maji}}\ and\ \bibinfo {author} {\bibfnamefont {M.~P.}\ \bibnamefont {Nowak}},\ }\bibfield  {title} {\bibinfo {title} {Scanning gate microscopy of nonretracing electron-hole trajectories in a normal-superconductor junction - code},\ }\bibfield  {journal} {\bibinfo  {journal} {Zenodo}\ }\href {https://doi.org/10.5281/zenodo.8409520} {10.5281/zenodo.8409520} (\bibinfo {year} {2023})\BibitemShut {NoStop}%
\bibitem [{QPC()}]{QPC_note}%
  \BibitemOpen
  \href@noop {} {}\bibinfo {note} {To set the working point of the QPC we close the othervise open boundary conditions at the edges opf the system along the $x$ direction.}\BibitemShut {Stop}%
\bibitem [{k_n()}]{k_note}%
  \BibitemOpen
  \href@noop {} {}\bibinfo {note} {For the discretized Hamiltonian used for the numerical calculation we use the equivalent formula for the wave-vector: $k_{e/h}=\frac{1}{a}\cos^{-1}({1-(\mu\pm E)/2t})$ with $t = \frac{\hbar^2}{2ma^2}$, where $a = 10$ nm is the discretization constant.}\BibitemShut {Stop}%
\bibitem [{\citenamefont {LeRoy}\ \emph {et~al.}(2002{\natexlab{b}})\citenamefont {LeRoy}, \citenamefont {Bleszynski}, \citenamefont {Topinka}, \citenamefont {Westervelt}, \citenamefont {Shaw}, \citenamefont {Heller}, \citenamefont {Maranowski},\ and\ \citenamefont {Gossard}}]{leroy2002imaging}%
  \BibitemOpen
  \bibfield  {author} {\bibinfo {author} {\bibfnamefont {B.~J.}\ \bibnamefont {LeRoy}}, \bibinfo {author} {\bibfnamefont {A.~C.}\ \bibnamefont {Bleszynski}}, \bibinfo {author} {\bibfnamefont {M.~A.}\ \bibnamefont {Topinka}}, \bibinfo {author} {\bibfnamefont {R.~M.}\ \bibnamefont {Westervelt}}, \bibinfo {author} {\bibfnamefont {S.~E.~J.}\ \bibnamefont {Shaw}}, \bibinfo {author} {\bibfnamefont {E.~J.}\ \bibnamefont {Heller}}, \bibinfo {author} {\bibfnamefont {K.~D.}\ \bibnamefont {Maranowski}},\ and\ \bibinfo {author} {\bibfnamefont {A.~C.}\ \bibnamefont {Gossard}},\ }\href@noop {} {\bibinfo {title} {Imaging coherent electron flow}} (\bibinfo {year} {2002}{\natexlab{b}}),\ \Eprint {https://arxiv.org/abs/cond-mat/0208194} {arXiv:cond-mat/0208194 [cond-mat.mes-hall]} \BibitemShut {NoStop}%
\end{thebibliography}%
\end{document}